\documentclass[pre,twocolumn,aps,superscriptaddress]{revtex4-2}
\usepackage[T1]{fontenc}
\usepackage[latin9]{inputenc}
\usepackage{array}
\usepackage{mathrsfs}
\usepackage{multirow}
\usepackage{amsmath}
\usepackage{amsthm}
\DeclareMathOperator{\sgn}{sgn}
\usepackage{adjustbox}
\usepackage{amssymb}
\usepackage{mathbbol}
\usepackage{graphicx}
\usepackage{latexsym}
\usepackage{textcomp}
\usepackage{mathtools}
\usepackage{xcolor}
\usepackage[citecolor=magenta,colorlinks=true]{hyperref}
\usepackage{bm}
\definecolor{mycolor1}{rgb}{0.1, 0.6, 0.6}

\begin{document}
\title{First contact breaking distributions in strained disordered crystals}
\author{Roshan Maharana}
\email{roshanm@tifrh.res.in}
\affiliation{Centre for Interdisciplinary Sciences, Tata Institute of Fundamental Research, Hyderabad 500107, India}

\author{Jishnu N. Nampoothiri}
\email{jishnu@brandeis.edu}
\affiliation{Martin Fisher School of Physics, Brandeis University, Waltham, MA 02454 USA}
\affiliation{Centre for Interdisciplinary Sciences, Tata Institute of Fundamental Research, Hyderabad 500107, India}

\author{Kabir Ramola}
\email{kramola@tifrh.res.in}
\affiliation{Centre for Interdisciplinary Sciences, Tata Institute of Fundamental Research, Hyderabad 500107, India}

\date{\today}

\begin{abstract}
We derive exact probability distributions for the strain ($\epsilon$) at which the first stress drop event occurs in uniformly strained disordered crystals, with quenched disorder introduced through polydispersity in particle sizes. We characterize these first stress drop events numerically as well as theoretically, and identify them with the first contact breaking event in the system. Our theoretical results are corroborated with numerical simulations of quasistatic volumetric strain applied to disordered near-crystalline configurations of athermal soft particles. We develop a general technique to determine the {\it distribution} of strains at which the first stress drop events occur, through an exact mapping between the cumulative distribution of first contact breaking events and the volume of a convex polytope whose dimension is determined by the number of defects $N_d$ in the system. An exact numerical computation of this polytope volume for systems with small numbers of defects displays a remarkable match with the distribution of strains generated through direct numerical simulations. Finally, we derive the distribution of strains at which the first stress drop occurs, assuming that individual contact breaking events are uncorrelated, which accurately reproduces distributions obtained from direct numerical simulations.
\end{abstract}

\maketitle

\section{Introduction}
Jammed materials arise in several natural contexts and have been studied in detail over the last two decades using a variety of theoretical, numerical and experimental techniques~\cite{o2002random,torquato2010jammed,behringer2018physics,van2009jamming,jaeger1996granular}. Such systems represent an extreme out of equilibrium scenario where temperature only plays a weak role in determining macroscopic properties~\cite{o2003jamming}. The mechanical properties of such disordered athermal materials have been of considerable interest~\cite{wyart2005rigidity,bi2011jamming,majmudar2005contact}. Although coarse-grained descriptions of such systems exist~\cite{lois2009stress,henkes2009statistical,yin2017modeling,hinkle2017coarse} and descriptions of the mechanical properties of jammed systems using the analogues of concepts such as entropy~\cite{henkes2007entropy}, scaling~\cite{goodrich2012finite,ramola2017scaling} and criticality~\cite{henkes2005jamming} have been developed to describe the statistical physics of athermal materials~\cite{bi2015statistical}, exact predictions starting from microscopic interactions as can be done for canonical thermal crystalline systems, are presently not available.

A well-known aspect of the mechanical properties of athermal amorphous systems is their anomalous rigidity, with failure governed by a {\it distribution} of external strain or shear ~\cite{wyart2005rigidity,cates1998jamming,karmakar2010predicting}. Several disordered systems can be characterized by their propensity to failure, which can be induced by the breaking of contacts between particles. 
In this regard, the first stress drop distribution provides insight into the stability and fragility of the phase of such a system~\cite{muller2015marginal,liu2010jamming}. Crucially, these quantities are also able to distinguish between the different regimes of stability of amorphous solids~\cite{hentschel2015stochastic,karmakar2010statistical}. However, the exact nature of these distributions and theoretical computations describing their dependence on microscopic interactions between the constituent particles remains unclear. It is therefore important to develop exactly solvable model systems akin to those in thermal statistical physics~\cite{baxter2016exactly}, to characterize the mechanical response and stability to external perturbations of disordered athermal systems.

In this work, we provide a step in this direction, making predictions for the distribution of the first stress drop in response to external strain for a near-crystalline athermal system, which is caused by a contact breaking event in the system. We consider an interesting example of a jammed near-crystalline system where several exact predictions of correlations and response can be made~\cite{acharya2020athermal,acharya2021disorder,das2021displacement,das2021long}. This system has allowed for the exact computation of many quantities of interest such as displacement fields in response to defects, interaction energies between defects \cite{acharya2021emergent}, as well as correlation functions \cite{das2021displacement}, starting from microscopic interactions. Therefore, the disordered crystal system is an apt candidate for a `model' system with which to understand the mechanical properties of amorphous athermal solids, such as first stress drop statistics. Since such distributions are capable of providing insights into the stability and fragility of amorphous solids, it is interesting to compute them exactly for the disordered crystal system.

In this study, we focus on the probability distributions of volumetric strains at which the first contact breaking events occur in an isotropically compressed disordered crystal system. We obtain the distributions of these events from both theoretical computations as well as numerical simulations and find an exact match between the two. We focus on volumetric strain instead of shear strain, as the computation of volumetric strain is more easily accessible theoretically. However, the formalism developed in this paper can also be extended to compute distributions of contact breaking events under the application of shear strain. 


The main result of this paper is to reduce the problem of computing the distribution of strain at which the first contact breaking occurs in a system with $N_d$ defects, to the computation of volumes of $N_d$-dimensional convex polytopes. The computation of convex polytope volumes is a well known problem in Mathematics and Computer Science, with a wealth of techniques developed over the years to address this problem, in particular due to its importance in the field of Linear Programming~\cite{sierksma2015linear}. This mapping therefore, allows an exact computation of the first contact breaking distribution in such systems using techniques developed to compute convex polytope volumes.

The outline of this paper is as follows. In Section~\ref{sec:disordered_crystal}, we introduce the disordered crystal system and detail the theoretical setup for computing the distributions of interest. We also identify the first plastic event with the first contact breaking event in this system and characterize these events both numerically as well as theoretically. In Section~\ref{sec:configuration space}, we develop a general technique to determine the {\it distribution} of strains at which the first contact breaking events occur, by utilizing the fact that the increase in strain or polydispersity generates a linear motion of constraints in the configuration space of disorder (see Fig.~\ref{fig_condition_surface}). In Section~\ref{sec:global distributions}, we provide an exact mapping of the computation of the cumulative distribution of the first contact breaking strain to the computation of the volume of an $N_d$-dimensional convex polytope. In Section~\ref{sec:local and global}, we focus on two specific types of plastic failures, local and global. In the former, we derive the distribution of volumetric strain required for a particular contact to break in the system. We then use this to compute a global distribution: the probability of the first contact breaking in the system at a given strain. We also show that local uncorrelated distributions of contact breaking provide an accurate description of first contact breaking events in such near-crystalline athermal systems. Finally, in Section~\ref{sec:scaling}, we discuss the behaviour of the first contact breaking distributions as the size of the system is increased, and show that they display an interesting logarithmic scaling with system size.

\section{Disordered Crystals Under Volumetric Strain}
\label{sec:disordered_crystal}


We consider a system of frictionless disks under isotropic compression in two dimensions, interacting through a one-sided pairwise potential of the form
\begin{align}
\nonumber
V_{ij} &=\frac{k}{\alpha}\left(1-\frac{\left|\vec{r}_{i j}\right|}{\sigma_{i j}}\right)^{\alpha} & \text { for } \left|\vec{r}_{i j}\right|<\sigma_{i j},\\
&=0 & \text { for } \left|\vec{r}_{i j}\right| \geq \sigma_{i j}.
\label{eq_potential}
\end{align}
Here $|\vec{r}_{ij}| = |\vec{r}_{i} - \vec{r}_{j}|$ is the distance between particles $i$ and $j$, located at positions $\vec{r}_{i}$ and $\vec{r}_{j}$ respectively and $\sigma_{ij}$ = $\sigma_i + \sigma_j$, is the sum of the radii of the two particles. The stiffness of the particles is quantified by the parameter $k$. 
The interparticle forces can be determined as
\begin{equation}
\vec{f}_{ij} = \frac{k}{\sigma_{ij}} \left(1- \frac{| \vec{r}_{ij}|}{\sigma_{ij}}\right)^{\alpha-1} \hat{r}_{ij},
\label{supp_force_law_equation}
\end{equation}
where $\hat{r}_{ij}$ is the unit vector along the line joining the centers of the $i^{th}$ and $j^{th}$ particle. 
For convenience we set $k = 1$.

Due to the one sided nature of the interaction, mechanical forces exist only when the system is overcompressed. In the present study, we choose $\alpha=2$, which corresponds to a harmonic pairwise potential between the particles. However, the methods developed in this paper can be easily extended to Hertzian ($\alpha=\frac{5}{2}$) or Hernian ($\alpha=\frac{3}{2}$) interactions as well.
We consider configurations at mechanical equilibrium, i.e. at energy minima, and perform quasi-static volumetric strain, generating perturbations that maintain the constraints of mechanical equilibrium.

\begin{figure}[t!]
\centering
\includegraphics[width=0.8\linewidth]{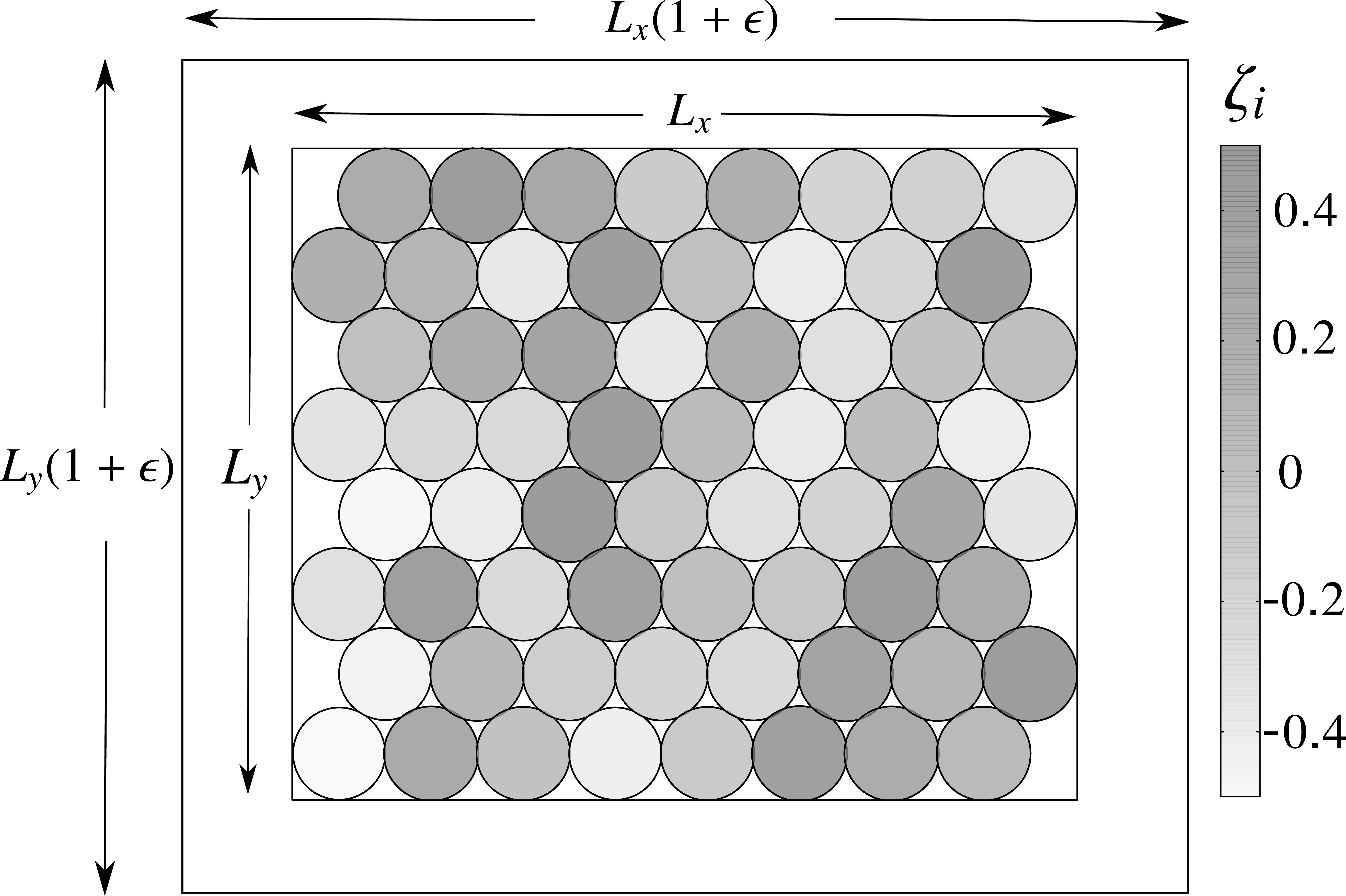}
\caption{
A disordered crystal configuration of a system of $64$ particles, generated with a realization of the quenched disorder $\{\zeta_i\}$, with disorder introduced in the radius of each particle $\sigma_i=\sigma_0(1+\eta \zeta_i)$. We study the stability of such a system to changes in the volume of the confining box under uniform volumetric strain, $[L_x,L_y] \rightarrow [L_x(1+\epsilon),L_y(1+\epsilon)]$, in the regime of small disorder strength $\eta$.
}
\label{fig_strain_schematic}
\end{figure}

We consider a system of $N$ particles confined in a commensurate box with linear dimensions $L_x$ and $L_y$ under periodic boundary conditions. The packing fraction of the system is given by
\begin{eqnarray}
\phi = \frac{\sum_{i=1}^{N} \pi \sigma_i^2}{L_x L_y},
\end{eqnarray}
To generate a disordered near-crystalline packing, we begin with an overcompressed triangular lattice of equal sized particles i.e. the packing fraction $\phi$ is greater than that of hard particles arranged in a triangular lattice ($\phi_c = \frac{\pi}{\sqrt{12}} \approx 0.9069$). The separation between two neighbouring particles in such an overcompressed system is given by $R_0 = 2\sigma_0\sqrt{\frac{\phi_c}{\phi}}$ . It is also convenient to define the rescaled lattice distance
\begin{equation}
\tilde{R}_0=\left(\frac{R_0}{2\sigma_0}\right)=\sqrt{\frac{\phi_c}{\phi}}.
\label{eq_rescaled_lattice}
\end{equation}
When $\tilde{R}_0 < 1$, the particles in the pure crystalline state (without the introduction of disorder) are in contact and the system is overcompressed.

Next, we create a disordered configuration by introducing quenched disorder in the particle radii ${\sigma_i}$ as
\begin{equation}
 \sigma_i=\sigma_0(1+\eta \zeta_i).
 \label{eq:disorder_radii_relation}
\end{equation}
Here $\{\zeta_i\}$ represents the quenched disorder in the system, with each $\zeta_i$ drawn from a uniform distribution between $-\frac{1}{2}$ and $\frac{1}{2}$ \cite{tong2015crystals}. The strength of the disorder is controlled by the polydispersity parameter $\eta$ (see Fig.~\ref{fig_strain_schematic}). We consider situations with disorder in the radii  restricted to a fixed subset of $N_d$ particles, with $N_d = 1,2, \hdots N$. To study the mechanical properties of this system, we apply a uniform volumetric strain to the system by changing the linear dimensions of the confining box as $L \to L(1+\epsilon)$, as depicted in Fig.~\ref{fig_strain_schematic}. 

Such athermal crystals with particle size polydispersity display an interesting transition to an amorphous phase at a critical value of the polydispersity for a given packing fraction. This transition is characterized by a diverging susceptibility to contact breaking across disorder realizations~\cite{tong2015crystals}, and has been termed a hidden order transition as the individual configurations do not display such a diverging susceptibility.

\subsection{Numerical Simulations~\label{sec:numerics}}

In our numerical simulations, we use the FIRE algorithm~\cite{bitzek2006structural} to generate energy minimized configurations. We consider an initial packing fraction of $\phi = 0.94$. At this packing fraction, the transition occurs at a critical value of the polydispersity $\eta_c \approx 0.04$ \cite{tong2015crystals}. In our study, we focus on the mechanical properties of this system and in particular, the first contact breaking events with increasing strain. We find that in this system the first plastic event coincides with the first contact breaking event, as we discuss in detail below. Other studies have also investigated the variation of contact number with pressure and polydispersity for weakly disordered crystals~\cite{ikeda2020jamming}. 

For a given disordered crystal packing characterized by a particular realization of the set of quenched random variables $\{\zeta_i\}$,  we uniformly tune either $\eta$ or $\epsilon$ starting from zero, and identify the value of $\eta$ or $\epsilon$ at which the first contact breaks. The procedure we follow is to start with a given disordered crystal configuration, increase $\eta$ or $\epsilon$ by a small amount and then perform energy minimization to attain mechanical equilibrium. We then calculate the number of neighbours for each particle in this energy minimized state. This procedure is repeated to determine the value of $\eta$ or $\epsilon$ at which the first contact breaking event occurs. The distributions of $\eta$ or $\epsilon$ where the first contact breaking events occur is obtained by repeating this process for multiple realizations of the quenched disorder $\{\zeta_i\}$. We also perform these simulations for different system sizes ranging from $N=16$ to $N=1024$ and a range of polydispersities $\eta \in [0.015,0.030] < \eta_c$. In this study, we focus specifically on the behavior of the system in the small polydispersity limit. Therefore, our study probes the regime of small disorder strength $\eta \approx 0.015$, which is well below the limit where the transition to an amorphous state is expected to occur~\cite{tong2015crystals}.


\subsection{Contact Breaking and Stress Drops~\label{sec:first_contact_plastic}}

\begin{figure*}[t!]
\centering
\includegraphics[width=0.8\linewidth]{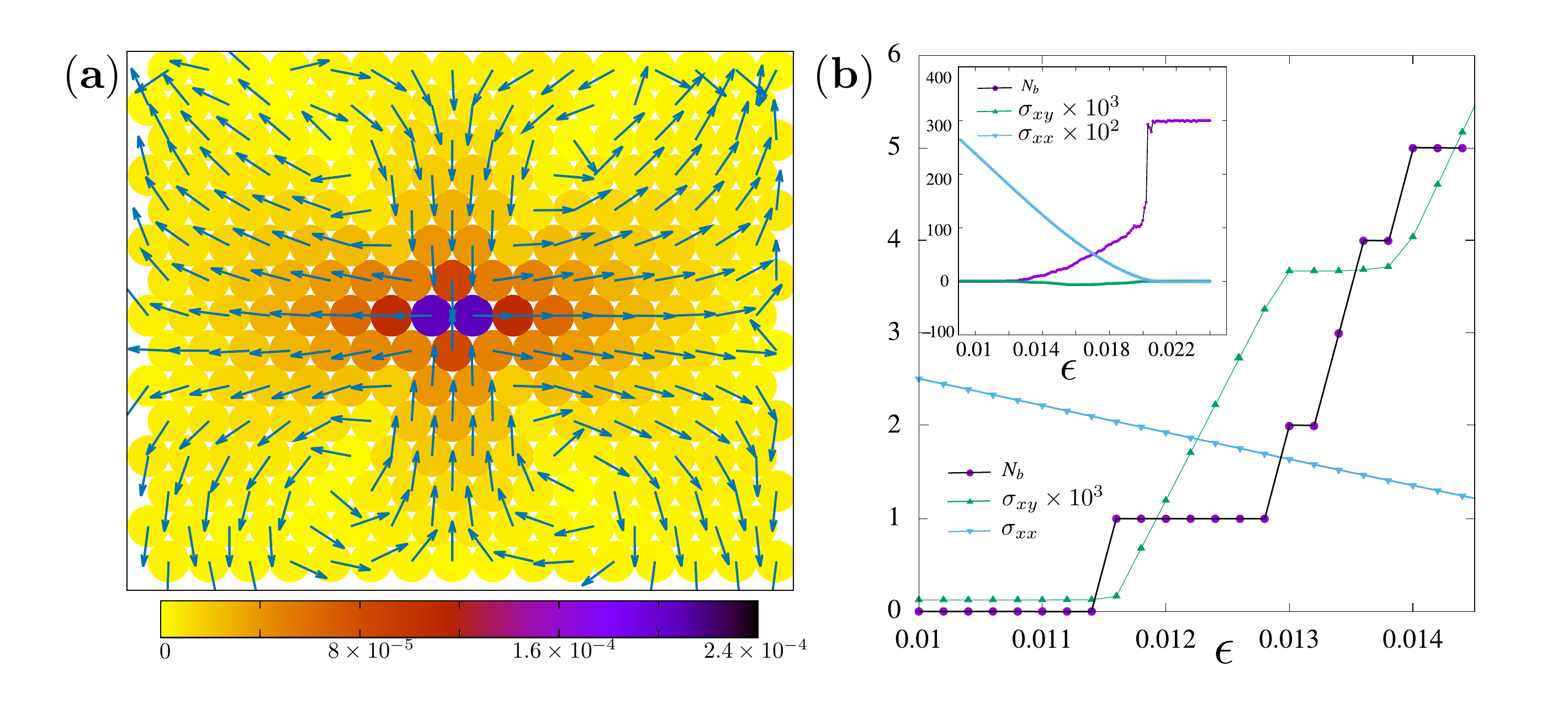}
\caption{\textbf{(a)} The change in the displacement field of each grain just before and after a contact breaking event. The system consists of $N=256$ grains at an initial packing fraction $\phi=0.94$, with polydispersity $\eta=0.015$. As a contact breaks in the system, there is a discontinuous change in the positions, centred on the broken contact. The color gradient represents the magnitude of the change in displacement fields relative to the particle size, whereas the direction of displacements are represented by unit arrows. \textbf{(b)} Change in the components of the global stress ($\sigma_{xx}$ and $\sigma_{xy}$) with increasing isotropic strain ($\epsilon$) for a particular realization of the quenched disorder $\{\zeta_i\}$. The system size considered is $N =100$. The number of broken contacts in the system $N_b$ increases discontinuously with strain, causing a discontinuous change in the global shear stress at each contact breaking event. The isotropic stress or the pressure ($\sigma_{xx}$) decreases continuously at the beginning of the contact breaking process. \textbf{(Inset)} The change in the components of the stress tensor, averaged over several realizations of disorder in particle sizes. Upon disorder average, $\sigma_{xx}$ decreases, while $\sigma_{xy}$ displays no significant change.
}
\label{stress_drop}
\end{figure*}

Earlier studies have identified discontinuous changes in system properties at contact breaking events in near-crystalline athermal systems~\cite{tuckman2020contact}. Similarly, the first contact breaking event with increasing strain leads to discontinuous changes in the displacement fields as well as the stresses of the system. In Fig.~\ref{stress_drop} {\bf (a)} we display the change in the displacement field of each grain just before and after a contact breaking event. As a contact breaks, there is a discontinuous change in the positions of the grains. These displacements form a quadrupolar pattern centred on the broken contact, reminiscent of Eshelby events in sheared amorphous materials \cite{lemaitre2009rate}.
In Fig.~\ref{stress_drop} {\bf (b)}, we display the global shear stress ($\sigma_{xy}$) of the system. This changes abruptly after the system undergoes the first contact breaking event with increasing volumetric strain.
Although for a given realization of the quenched disorder $\sigma_{xy}$ displays an abrupt change, it displays no significant change when {\it averaged} over several realizations of the disorder.
On the other hand, upon disorder average, $\sigma_{xx}$ shows a continuous decrease as the volumetric strain is increased.
The change in the components of the stress tensor, averaged over the several realizations of the disorder are displayed in the inset of Fig.~\ref{stress_drop} {\bf (b)}.

\begin{figure}[t!]
\centering
\includegraphics[width=1.00\linewidth]{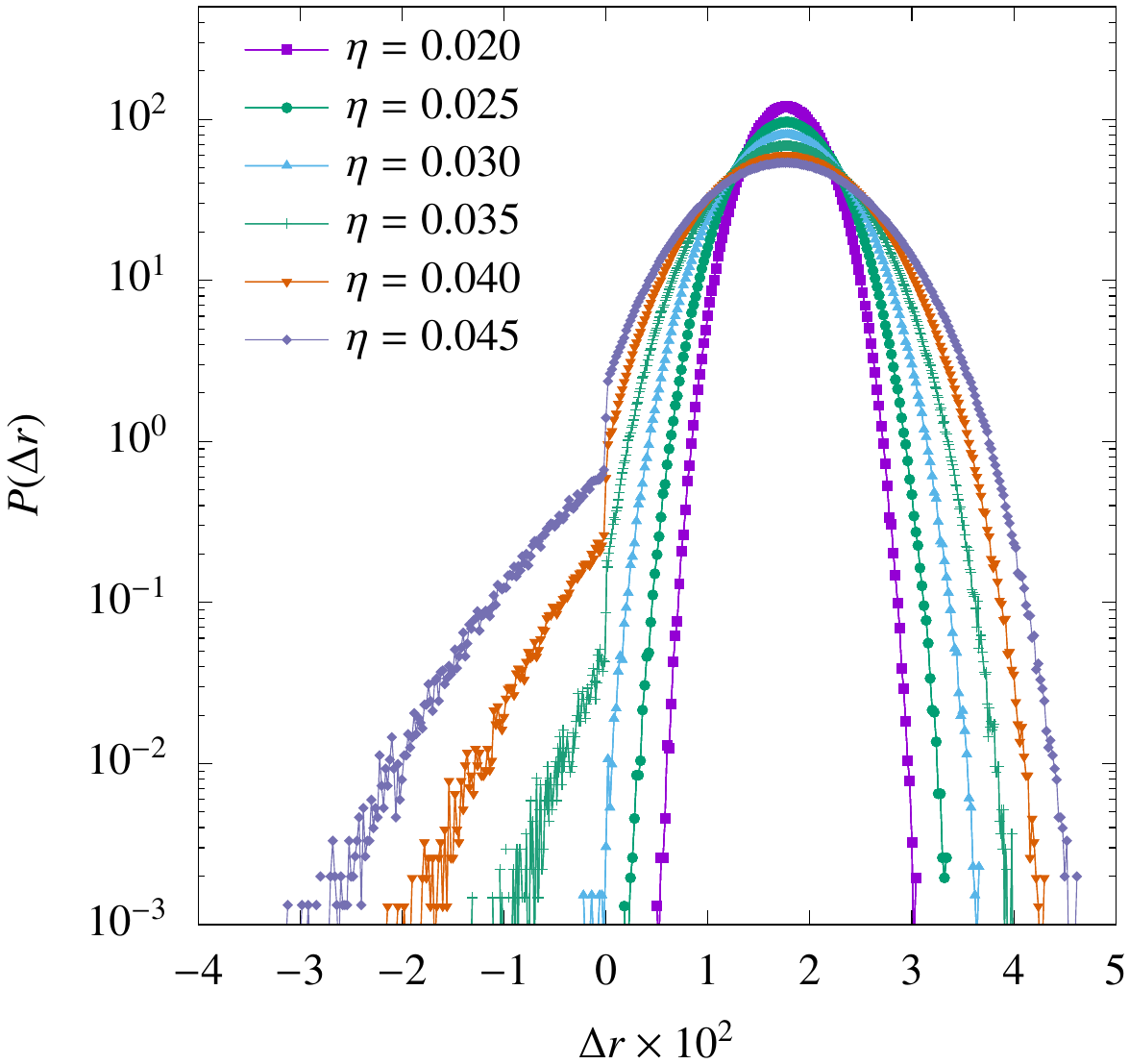}
\caption{
Distribution of overlap distances $\Delta r = \sigma_i+\sigma_{j}-|\vec{r}_{i}-\vec{r}_j|$ for pairs of nearest neighbour particles $i$, $j$, with increasing polydispersity $\eta$. The region with $\Delta r < 0$ represents broken contacts between the particles. The discontinuity at $\Delta r=0$ points to the fact that the contact breaking event produces a discontinuous change in the displacement fields. The distribution displayed is for a system of size $N=256$ at zero strain.
}
\label{fig_deltar_distribution}
\end{figure}

In Fig.~\ref{fig_deltar_distribution}, we plot the distribution of the gaps between neighbouring particles $\Delta r = \sigma_i+\sigma_{j}-|\vec{r}_{i}-\vec{r}_j|$, which displays a discontinuity at $\Delta r=0$. As $\Delta r< 0$ represents broken contacts, this distribution makes it clear that once a contact breaking event occurs, the corresponding particles move finite distances away, as the system settles into a {\it new} energy minimum. Therefore a linearized perturbation analysis of all the particles with respect to polydispersity or volumetric strain is no longer valid after the system encounters a contact breaking event. This leads to the discontinuous change in the stress distribution. 

\section{Linearized trajectories in configuration space~\label{sec:configuration space}}

We begin our theoretical analysis by studying {\it energy minimized} configurations of the system generated as a response to the microscopic disorder. We use the theory developed recently in Refs.~\cite{acharya2020athermal,das2021long,acharya2021disorder}, which allows an exact determination of the {\it perturbed} configuration as a response to the disorder. This provides a one to one map between a given realization of the quenched disorder $\{ \zeta_i\}$ and the {\it position} of the system in the configuration space of particle positions. As a consequence of this one to one mapping, we can then follow the trajectory of the system through configuration space as the polydispersity $\eta$ or the strain $\epsilon$ is increased. The direction of this motion is determined by the quenched random variables $\{ \zeta_i \}$ as we show below. The contact breaking process can therefore be naturally mapped onto the process of this trajectory encountering a boundary defined by a contact breaking condition in configuration space.

The volumetric strain can be equivalently implemented by keeping the volume of the box fixed while changing the size of each particle as $\sigma_i \rightarrow \frac{\sigma_i}{1+\epsilon}$. This change in volume changes the packing fraction $\phi$ as
\begin{equation}
\phi \rightarrow \frac{\phi}{(1+\epsilon)^2}~.
\end{equation}
The quenched change in the radii at each site $\vec{r} \equiv i$ representing the microscopic disorder in the system, can equivalently be represented as
\begin{equation}
\delta \sigma(\vec{r})= -\left(\frac{\epsilon}{1+\epsilon}\right)\sigma_0  +\left(\frac{1}{1+\epsilon}\right) \sigma_0\eta \zeta(\vec{r}).
\label{eq_delsigma}
\end{equation}
We note that the two protocols, increasing the volume of the system or decreasing the sizes of the particles, produce the same (scaled) displacement fields. However, they produce different changes in the energy of the system. However, for the relative displacements between particles, there exists a one to one map between the protocols.

\subsection{Linearized Displacement Fields}

We next detail a procedure to obtain the exact displacement fields generated in response to the introduction of small disorder in particle sizes, as developed in Refs.~\cite{acharya2020athermal,das2021long,acharya2021disorder}. As the starting and ending configurations both satisfy the conditions of mechanical equilibrium, we have $\sum_{j} f^{x}_{ij} = 0$, and $\sum_{j} f^{y}_{ij} = 0$, for each particle at site $i \equiv \vec{r}$. 
Here $f^{x}_{ij}$, $f^{y}_{ij}$ are the components of the inter-particle contact force between particles $i$ and $j$.
To determine the displacement fields, {\it all} the force balance equations must be simultaneously satisfied, which through the force-law in Eq.~(\ref{supp_force_law_equation}) yields a unique solution for particle displacements \cite{acharya2020athermal, acharya2021disorder}. As there are $2 N$ displacement variables $\{\delta x_i, \delta y_i\}$, and $2 N$ equations of force balance, this provides enough equations to determine the displacement fields, given the microscopic disorder $\{ \delta \sigma \}$.

However, the force law in Eq.~(\ref{eq_potential}) is non-linear in the displacement fields $\{ \delta x, \delta y \}$.
This non-linearity can be circumvented with a systematic perturbation expansion about the crystalline ordered state, to linear order as well as higher orders \cite{acharya2020athermal,das2021long,acharya2021disorder}. With the introduction of disorder in the particle sizes, the positions of the particles deviate from their crystalline values $\{ \vec{r}_{i}^{(0)} \} = \{x_{i}^{(0)}, y_{i}^{(0)}\}$, to new positions $\{  \vec{r}_{i}^{(0)} + \delta \vec{r}_i \} = \{x_{i}^{(0)} + \delta x_i, y_{i}^{(0)}+\delta y_i\}$ which satisfy the constraints of mechanical equilibrium. Next, the displacements fields can be expressed as an expansion in the strength of the disorder as
\begin{eqnarray}
\delta \vec{r}_i = \delta \vec{r}^{(1)}_{i} + \delta \vec{r}^{(2)}_{i} + \delta \vec{r}^{(3)}_{i} + \hdots.
\label{displacement_eq}
\end{eqnarray}
Here $\{ \delta \vec{r}^{(n)}_{i} \} =  \{ \delta x^{(n)}_{i}, \delta y^{(n)}_{i} \}$ represent the correction to the displacement fields of magnitude $\mathcal{O}(\eta^n)$. As the coefficients in the perturbation expansion only depend on the initial crystalline structure, the force balance equations can be solved hierarchically at every order using Fourier transforms to obtain the displacement fields~\cite{acharya2021disorder}. The change in radii $\{ \delta \sigma \}$ and the displacement fields at lower order $\{ \delta \vec{r}^{(n)}_i \}$ can be interpreted as sources that generate the displacement fields at higher orders $\{ \delta \vec{r}^{(n+1)}_i \}$. This allows us to construct a one-to-one map between the disorder in the particle radii and the displacement of each particle from the crystalline position. In this work, we restrict ourselves to the linear order solutions and this linear order approximation yields contact breaking distributions which match the observed distributions from numerical simulations with sufficient accuracy. Therefore, we can conclude that for small disorder, the linearized approximation provides accurate descriptions of the displacement fields up to the first contact breaking event.

At linear order, the displacements in response to the  disorder $ \{ \delta \sigma_{i} \}$ can be expressed in Fourier space as
\begin{eqnarray}
\delta \tilde{ r}^{\mu,(1)}(\vec{k}) = \tilde{G}^{\mu}(\vec{k})\delta \tilde{\sigma}(\vec{k}).
\end{eqnarray}
Here $ \delta \tilde{\sigma}(\vec{k}) = \sum_{\vec{k}} e^{i\vec{k}\cdot \vec{r}} \delta \sigma(\vec{r})$ and  $\delta r^{\mu}$ refers to $\delta x$ and $\delta y$ for $\mu = x,y$ respectively. $\delta \tilde{r}^{\mu}(\vec{k})$ and   $\tilde{G}^{\mu}(\vec{k})$ represent the Fourier transforms of the displacement fields and Green's functions. As the system is a perfect triangular lattice before the introduction of disorder, these functions are non-zero at the reciprocal lattice vectors of the triangular lattice, $\vec{k} \equiv (k_x,k_y) \equiv \left( \frac{2 \pi l}{2 L}, \frac{2 \pi m}{L} \right)$ \cite{horiguchi1972lattice}. We have provided the exact expressions for these Green's functions in Appendix~\ref{appendix:A}. Therefore, to linear order in the perturbation expansion, the $x$ and $y$ components of the displacement of a particle situated at $\vec{r}$ can be expressed as
\begin{equation}
    \begin{aligned}
        &\delta x(\vec{r}) = \sum_{\vec{r}'} G^{x}(\vec{r}-\vec{r}')\delta \sigma(\vec{r}'),\\
        &\delta y(\vec{r})=\sum_{\vec{r}'} G^{y}(\vec{r}-\vec{r}')\delta \sigma(\vec{r}'),
    \end{aligned}
    \label{delr_def}
\end{equation}
Here, the displacements are expressed in terms of the Green's functions that relate the displacements at site $\vec{r}$ to the disorder in the particle radii at site $\vec{r}'$. Next, we use the following property of the Green's functions
 \begin{equation}
    \sum_{\vec{r}} G^{x}(\vec{r})=\sum_{\vec{r}} G^{y}(\vec{r})=0,
    \label{Green_prop}
 \end{equation}
to show that the first term in the source in Eq.~(\ref{eq_delsigma}) does not contribute to the change in the displacement fields. Therefore, using Eqs.~\eqref{eq_delsigma}, \eqref{delr_def} and \eqref{Green_prop} we arrive at the displacement fields as a response to the quenched disorder and external strain
\begin{equation}
    \begin{aligned}
        &\delta x(\vec{r})=\frac{\sigma_0\eta}{1+\epsilon} \sum_{\vec{r}'} G^{x}(\vec{r}-\vec{r}') \zeta(\vec{r}'),\\
        &\delta y(\vec{r})=\frac{\sigma_0\eta}{1+\epsilon} \sum_{\vec{r}'} G^{y}(\vec{r}-\vec{r}') \zeta(\vec{r}').
    \end{aligned}
    \label{eq_linearized_positions}
\end{equation}
We have therefore established a map between the volumetric strain ($\epsilon$) of the system and the displacement fields that satisfy the conditions of mechanical equilibrium through the linearized force law, for a given configuration of quenched radii. Within the above linear framework, which is valid for small $\eta$ and $\epsilon$, the displacement of each particle from its crystalline position is linear. This induces a linear trajectory of the system in the $2N$ dimensional phase space of particle coordinates $\{ x_i, y_i\}$. Therefore, the phase space trajectories under a change of $\eta$ or $\epsilon$ are straight lines whose slopes are determined by the initial quenched variables $\{\zeta_i\}$.

\subsection{Contact Breaking Conditions~\label{sec:contact breaking conditions}}



Having established a one-to-one map between the displacement fields $\{\delta x, \delta y\}$ and the underlying microscopic disorder $\{ \delta \sigma \}$, we can reformulate the contact breaking conditions directly in terms of the quenched disorder $\{\zeta_i\}$. We can therefore map each point in the configuration space of particle positions to a point in the phase space of the quenched disorder $\{\zeta_i\}$. The phase space of quenched disorder $\{\zeta_i\}$ is an $N_d$-dimensional hypercube defined by the range of the disorder variables $|\zeta_i| < \frac{1}{2}$. 

Next, the condition for a contact between the $i^{th}$ and $k^{th}$ particle to remain unbroken in the presence of disorder can be expressed as
\begin{equation}
\begin{aligned}
   \sum_{\mu} (r_i^{\mu(0)}-r_k^{\mu(0)}&+\delta r_i^{\mu}- \delta r_k^{\mu})^2<(2\sigma_0+\delta \sigma_{i}+\delta \sigma_k)^{2}.
\end{aligned}
\end{equation}
Now, in the limit of small disorder i.e. the displacement response to the disorder in the particle sizes being much smaller than the initial separation between particles, we can linearize the above equation as
\begin{equation}
\begin{aligned}
   R_0^2+\sum_{\mu=x,y}\left[ 2r_{ik}^{\mu(0)}\left(\delta r_{ik}^{\mu}\right)\right]<4\sigma_0^2+4\sigma_0\delta \sigma_{ik}.
\end{aligned}
\end{equation}
Without loss of generality we choose the particular contact to be the $j^{th}$ contact of the particle at position $i \equiv \vec{r}$ (see Fig.~\ref{fig_schematic2}). Using the linearized displacement fields given in Eq.~(\ref{eq_linearized_positions}) and the condition above, we can predict when the $j^{th}$ contact of the particle at $\vec{r}$ undergoes a contact breaking event for the first time as
\begin{equation}
\sum_{\vec{r}'}C_j(\vec{r},\vec{r}')\delta \sigma(\vec{r}') < 2\sigma_0\left(1- \tilde{R}_0^2\right),
\label{coefficients}
\end{equation}
where the volumetric strain appears through the scaling of the radii in Eq.~(\ref{eq_delsigma}) and $\tilde{R}_0$ represents the rescaled lattice distance defined in Eq.~(\ref{eq_rescaled_lattice}). The details of this derivation are provided in Appendix~\ref{sec:contact breaking conditions}.
As the above equations describe the conditions for contact breaking for the $j^{\mathrm{th}}$ bond at site $\vec{r}$, the total number of conditions is the total number of bonds in the system i.e. $3N$. Since we keep the particle sizes and the initial packing fraction fixed while performing the volumetric strain, the RHS of the above equation can be treated as a constant which only depends on the initial conditions. The coefficients $C_j(\vec{r},\vec{r}')$ relate the displacements that arise in the contact breaking condition at $\vec{r}$ from the local disorder at $\vec{r}'$. These are expressible in terms of Green's functions, which are exactly known. We have
\begin{align}
\label{coefficients2}
    &C_j(\vec{r},\vec{r}')=-2 \left( \delta_{\vec{r},\vec{r}'}+\delta_{\vec{r}+\vec{\Delta}_j,\vec{r}'} \right)\\
    & + 2\tilde{R}_0\left[ \cos\left(\frac{2\pi j}{6}\right)\left(G^{x}\left(\vec{r}'-(\vec{r}+\vec{\Delta}_j)\right)-G^{x}(\vec{r}'-\vec{r})\right) \right.\nonumber\\
    &\left. ~~~~~+ \sin\left(\frac{2\pi j}{6}\right)\left(G^{y}\left(\vec{r}'-(\vec{r}+\vec{\Delta}_j)\right)-G^{y}(\vec{r}'-\vec{r})\right)\right]\nonumber. 
\end{align}
Here $\vec{\Delta}_j$ correspond to the fundamental translation vectors of the triangular lattice which are represented in Fig.~\ref{fig_schematic2}. There are $6$ such vectors corresponding to the $6$ nearest neighbours. All the coefficients $C_j(\vec{r},\vec{r}')$ depend only on the underlying structure of the crystalline state at zero disorder and therefore are independent of parameters such as strain $\epsilon$ or the strength of disorder $\eta$. Due to the translation invariance of the system, the coefficients corresponding to any two contacts along the $j$-direction are equal, i.e.
\begin{equation}
C_j(\vec{r},\vec{r}')=  C_j(\vec{r}+\vec{r}_0,\vec{r}'+\vec{r}_0),
\end{equation}
for any translation by a distance $\vec{r}_0$ on the lattice.

\begin{figure}[t!]
\includegraphics[width=1.0\linewidth]{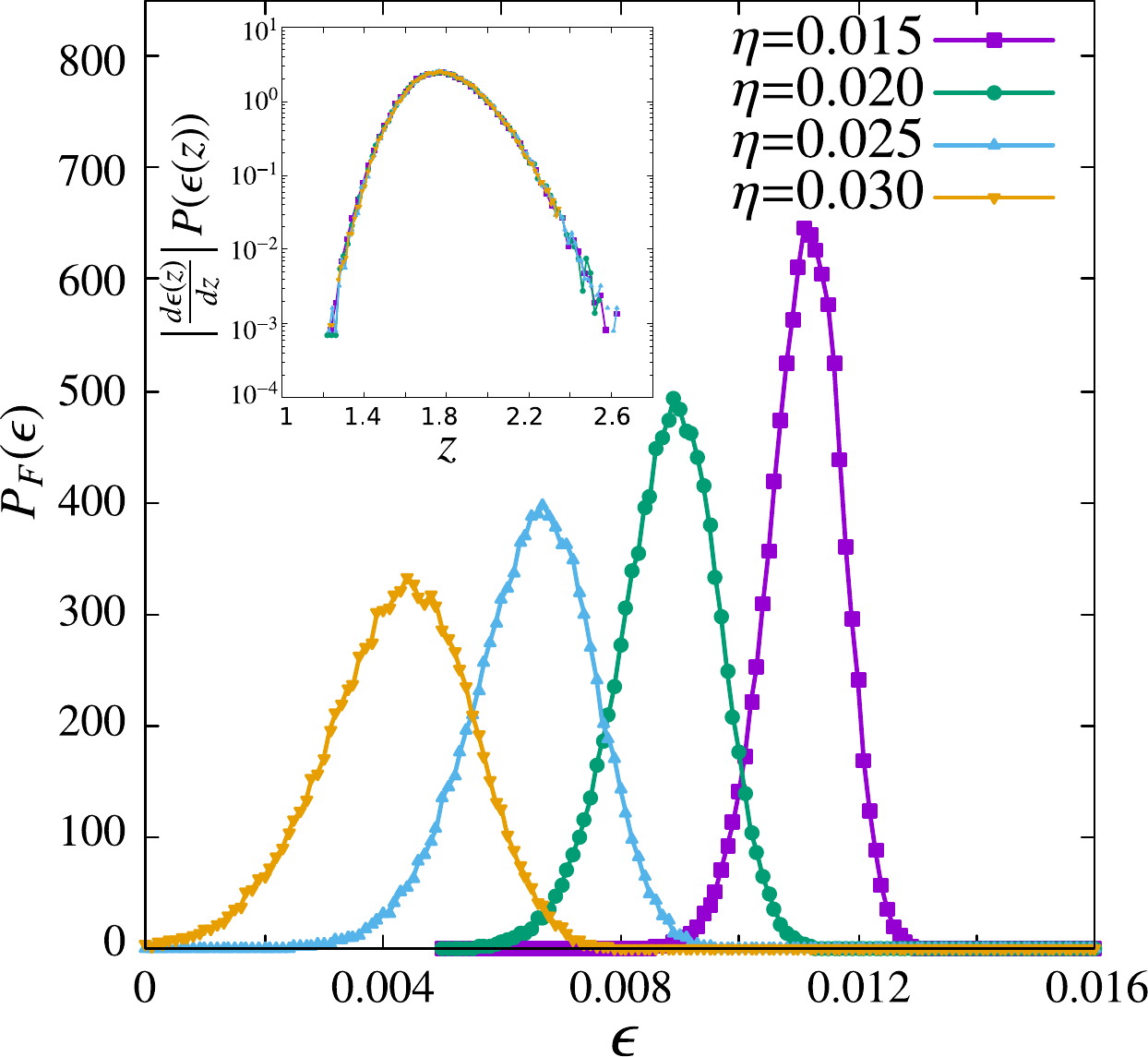}
\caption{Distribution of first contact breaking strains for a system of size $N=256$ at different values of polydispersity ($\eta$) obtained from direct numerical simulations. At $\eta=0$, i.e. for a perfectly ordered system of soft particles, the distribution is a delta function located at $\epsilon= \left(\frac{1-\tilde{R}_0^2}{1+\tilde{R}_0^2}\right)$ (for $\phi=0.94$, $\epsilon=0.0179$), at which all the contacts in the system break simultaneously. {\bf(Inset)} The same distributions plotted as a function of the variable $z(\eta,\epsilon) = \frac{2-2\tilde{R}_0^2}{\eta}\left(1-\epsilon \left(\frac{1+\tilde{R}_0^2}{1-\tilde{R}_0^2}\right)\right)$, displaying a near-perfect scaling collapse.
}
\label{fig_different_eta_256}
\end{figure}

Now, substituting the value of $\delta\sigma(\vec{r}')$ from Eq.~\eqref{eq_delsigma} in Eq.~\eqref{coefficients}, and making use of a property of these coefficients: $\sum_{\vec{r}'}C_j(\vec{r},\vec{r}')=-4$ since $\sum_{\vec{r}'}G^{\mu}(\vec{r} -\vec{r}')=0$. For the contact $j$ of the particle situated at $\vec{r}$, we obtain
\begin{equation}
\sum_{\vec{r}'}C_j(\vec{r},\vec{r}') \zeta(\vec{r}') =
\underbrace{
\frac{2-2\tilde{R}_0^2}{\eta}\left(1-\epsilon \left(\frac{1+\tilde{R}_0^2}{1-\tilde{R}_0^2}\right)\right)}_{z(\eta,\epsilon)}.
\label{eq_cb_condition}
\end{equation}
The form of the contact breaking conditions in the above equation immediately suggests the following scaling variable 
\begin{equation}
z(\eta,\epsilon) = \frac{2-2\tilde{R}_0^2}{\eta}\left(1-\epsilon \left(\frac{1+\tilde{R}_0^2}{1-\tilde{R}_0^2}\right)\right),
\label{eq:z_scaling_defn}
\end{equation}
along with the inversion
\begin{eqnarray}
\epsilon(z, \eta)=\left(\frac{1-\tilde{R}_0^2}{1+\tilde{R}_0^2}\right) - \frac{ \eta z}{2\left(1+\tilde{R}_0^2\right)}.
\end{eqnarray}
We note that this scaling does not depend on the exact form of the expressions in the LHS of Eq.~(\ref{eq_cb_condition}), and can be predicted simply by assuming the linear behaviour of the displacement fields and translation invariance.  As the LHS of Eq.~(\ref{eq_cb_condition}) is derived from the force law at a given packing fraction, it does not depend on other system parameters such as the strain, or the strength of the disorder. Therefore, for a given packing fraction, we may use this as a scaling variable for the contact breaking distributions. This suggests that the scaling should be valid for all contact breaking distributions, even in the presence of a finite number of contact breaking events.
The scaling of the first contact breaking strain distribution using the above scaling variable is displayed in the inset of Fig.~\ref{fig_different_eta_256}, producing a very good scaling collapse.

Interestingly, for a given realization of the quenched disorder $\{\zeta_i\}$, we can find the value of the strain up to which a {\it specific} contact represented by ($\vec{r},j$) does not break. We have
\begin{equation}
    \epsilon_{\vec{r},j}=  \left(\frac{1-\tilde{R}_0^2}{1+\tilde{R}_0^2}\right) - \frac{ \eta}{2\left(1+\tilde{R}_0^2\right)}\sum_{\vec{r}'}C_j(\vec{r},\vec{r}') \zeta(\vec{r}').
\end{equation}
For a system with a large number of particles, contacts may break even in the unstrained system as a result of the introduction of polydispersity. Therefore there exists a critical polydispersity for a given system size, up to which the contacts in the system remain unbroken. This relationship between the range of polydispersity and the system size for which no contacts are broken in the unstrained system is described in Appendix~\ref{threshold polydispersity}.

\section{First Contact Breaking distributions using Polytope Volumes~\label{sec:global distributions}}

We next derive the distribution of strains at which the {\it first} contact in the system is broken as a response to the quenched polydispersity and increasing volumetric strain. 
In order to compute this distribution, we first compute $F(\epsilon)$, the cumulative probability that none of the contacts in the system break up to a strain $\epsilon$. 
Having computed this local distribution, it is straightforward to extract the distribution of $\epsilon$ at which the first contact breaks. The change in this cumulative probability between the strains $[\epsilon, \epsilon + d \epsilon]$ represents the probability that the first contact in the system breaks within this interval. We, therefore have
\begin{equation}
 P_F(\epsilon)=-\frac{dF(\epsilon)}{d\epsilon}=\left|\frac{\partial F(z)}{\partial z} \frac{dz}{d\epsilon} \right|.
 \label{prob_vol}
\end{equation}
The cumulative probability $P_F$ represents the realizations of the quenched disorder that satisfy {\it all} possible contact breaking conditions. 

\begin{figure}[t!]
\centering
\includegraphics[width=0.90\linewidth]{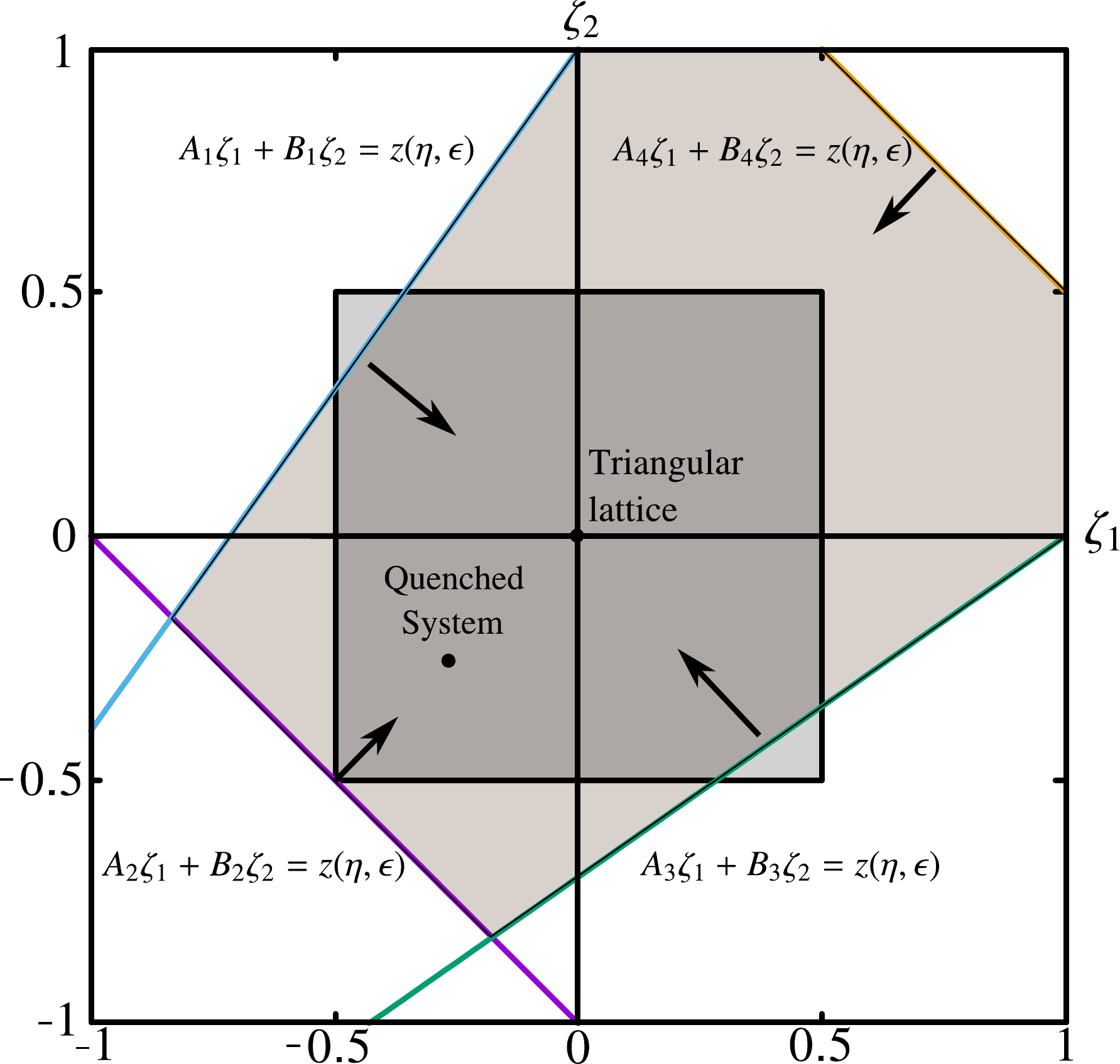}
\caption{
A schematic representation of contact breaking conditions in the phase space of quenched disorder $\{\zeta_i\}$. Here, disorder is introduced in the radii of two particles ($N_d=2$) making the phase space two dimensional. The arrows depict the motion of these conditions with increasing value of polydispersity $\eta$ or strain $\epsilon$. The origin corresponds to the initial ordered triangular lattice and the filled circle represents a particular realization of the quenched disorder $\{\zeta_i\}$. This system undergoes a contact breaking event for the first time when one of the conditions reaches the position $\{\zeta_i\}$ for the first time. Here, $z(\eta,\epsilon)$ is the scaling variable given in Eq.~\eqref{eq:z_scaling_defn}.}
\label{fig_condition_surface}
\end{figure}

Next, it is more convenient to consider the contact breaking conditions in the phase space spanned by the disorder variables. These represent the relevant variables in the system as the displacements can be derived from them. For $N_d$ particles with quenched disorder, this forms an $N_d$-dimensional space.  A volumetric strain of the system induces a motion of the contact breaking conditions in this $N_d$-dimensional phase space of quenched disorder, as illustrated in Fig.~\ref{fig_condition_surface}. Since the quenched disorder is uniformly distributed within the $N_d$-dimensional hypercube, the cumulative probability $F(\epsilon)$ can therefore be extracted as the volume of the polytope defined by these conditions, and the limits of the distribution $\zeta_{i} \in [-\frac{1}{2},\frac{1}{2}]$. Therefore, we are able to reduce the computation of the cumulative distribution of first contact breaking strains to computation of $N_d$-dimensional polytope volumes. The distribution of the first contact breaking strain can then be extracted using Eq.~\eqref{prob_vol}.

\begin{figure*}[t!]
\centering
\includegraphics[width=1.0\textwidth]{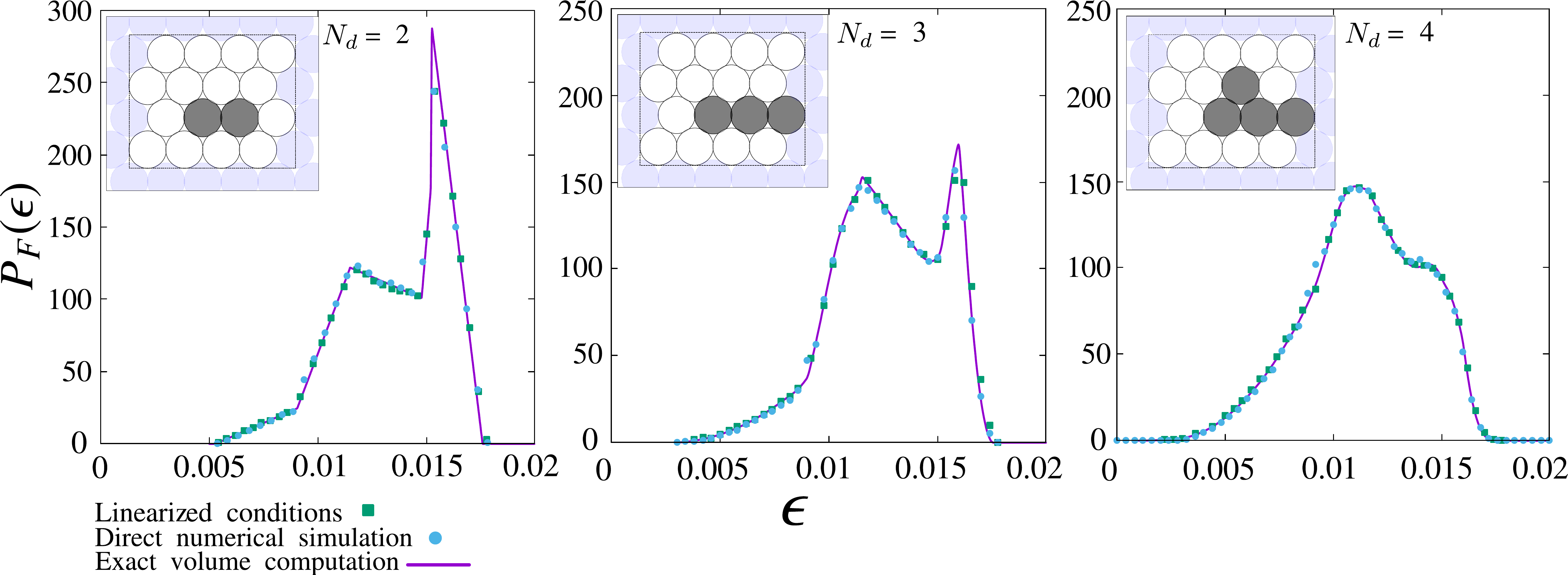}
\caption{
The distribution of strain at which the first contact breaking event occurs $P_F(\epsilon)$, computed using three different methods. (i) A Monte-Carlo sampling, utilizing the linearized contact breaking conditions in Eq.~\eqref{eq_cb_condition}.
(ii) An exact numerical computation of volume enclosed by the contact breaking conditions, using algorithms developed to compute convex polytope volumes and (iii) A direct numerical simulation of the system. The plots display the first contact breaking strain distributions as
a function of strain ($\epsilon$), for $N_d=2,3,4$ i.e. increasing number of particles with disorder in the radii. The results displayed are for a system size $N =16$. The {\bf insets} display a schematic of the system, with the darker particles representing the subset in which quenched disorder is introduced. The lighter shaded particles represent periodic copies of the system.}
\label{n_disorder}
\end{figure*}

To illustrate this procedure, we discuss a simple case where disorder is introduced in the radii of only two particles in the system. We choose particles $p$ and $q$ situated at $\vec{r}_p$ and $\vec{r}_q$ respectively. In this case, the quenched disorder $\zeta(\vec{r})=0 $ at all sites except $\zeta(\vec{r}_p) $ and $\zeta(\vec{r}_q) $. Then, the condition in Eq.~\eqref{eq_cb_condition} for a single contact ($\vec{r},j$) to break can be written as
\begin{equation}
\begin{aligned}
C_{j}(\vec{r},\vec{r}_p) \zeta(\vec{r}_p)+C_{j}(\vec{r},\vec{r}_q) \zeta(\vec{r}_q) = z(\eta,\epsilon),
\end{aligned}
\label{2particle}
\end{equation}
while the rest of the bond breaking conditions for contacts $(\vec{r}',j')$ are not violated, i.e.
\begin{equation}
\begin{aligned}
C_{j'}(\vec{r}',\vec{r}_p) \zeta(\vec{r}_p)+C_{j'}(\vec{r}',\vec{r}_q) \zeta(\vec{r}_q) < z(\eta,\epsilon),
\end{aligned}
\label{2particleb}
\end{equation}
{for $j \ne j'$.}
Using this illustrative example, with disorder in particle sizes at $\vec{r}_p$ and $\vec{r}_q$, the contact breaking conditions are a set of straight lines in the $\left[\zeta(\vec{r}_p),\zeta(\vec{r}_q)\right]$ plane. As the quenched disorder $\zeta(\vec{r}_p)$ and    $\zeta(\vec{r}_q)$ range from $[-\frac{1}{2},\frac{1}{2}]$, the convex volume enclosed by the contact breaking conditions in Eqs.~\eqref{2particle} and \eqref{2particleb} and the boundary lines $\zeta(\vec{r}_p)=\pm \frac{1}{2}$ and $\zeta(\vec{r}_q)=\pm \frac{1}{2}$ yields the probability that no contact is broken at this value of $\eta$ and $\epsilon$. 

This procedure can now be generalized to study bond breaking anywhere on the lattice as $(\vec{r},j)$ can represent any arbitrary contact amongst the $3N$ contacts present in the system. In this case, the contact breaking conditions dictating that all $3N$ contacts remain unbroken provide the cumulative distribution of the first contact breaking strain. The general form for the volume of the convex polytope enclosed by all the contact breaking conditions in the $N_d$-dimensional space of quenched disorder (with $N_d =N$) is given by
\begin{equation}
\begin{aligned}
    &F(\epsilon,\eta)=\\
    &\int_{-\frac{1}{2}}^{\frac{1}{2}}\prod_{\vec{r} ''} d \zeta(\vec{r}'') \prod_{(\vec{r},j)}\left[\Theta \left(z(\epsilon,\eta)-\sum_{\vec{r} '} C_j(\vec{r},\vec{r} ')\zeta(\vec{r} ')\right)\right].
    \end{aligned}
    \label{eq_exact_volume}
\end{equation}

For the case of disorder introduced in the radii of two particles, we can compute the volume of the convex polytope in two dimensions using Eqs.~\eqref{2particle} and~\eqref{eq_exact_volume}, and then using Eq.~\eqref{prob_vol} we can determine the probability distribution of first contact breaking.  Similarly, we can introduce disorder in $N_d = 3$, $4$ or $5$ particles. 
The dimension of the phase space of quenched disorder therefore also increases depending on the number particles in which disorder is introduced. In Fig.~\ref{n_disorder}, we plot the first contact breaking strain distribution obtained by computing the associated polytope volumes with increasing strain ($\epsilon$). In low dimensions: $N_d = 2,3$ and $4$, an exact computation of the polytope volume is feasible and we use the $lrslib$ package~\cite{avis2000revised} to compute these exact polytope volumes. The computation of exact convex polytope volumes become impractical at large dimensions~\cite{dyer1988complexity}. However, in recent years random approximation algorithms which are able to compute volumes of polytopes with theoretical efficiencies of up to $\mathcal{O}(n)$ have been developed. Therefore, the volume of the polytopes can be computed using random approximation algorithms~\cite{cousins2016practical}, to determine contact breaking distributions.


The infeasibility of exact convex polytope volume computations in higher dimensions necessitates the use of alternate methods to compute the first contact breaking strain distribution. In Fig.~\ref{n_disorder}, we present comparisons between the contact breaking strain distribution computed using three different methods:  (i) A Monte-Carlo sampling, utilizing the linearized contact breaking conditions in Eq.~\eqref{eq_cb_condition}.
(ii) An exact numerical computation of volume enclosed by the contact breaking conditions, using algorithms developed to compute convex polytope volumes and (iii) A direct numerical simulation of the system. We find that the distributions computed using these alternate procedures match well with the distribution predicted from the convex polytope volume computation and therefore, we employ these alternate methods to compute the contact breaking distributions in higher dimensions.

\section{First Contact Breaking Distributions using Local Conditions\label{sec:local and global}}

In this Section, we use the formulation developed above to compute the first contact breaking distribution in two steps: (i) we compute a `local' contact breaking distribution: namely the probability of a particular contact in the system breaking at a given strain, (ii) we use this local distribution to compute the distribution of strain at which the first contact breaking events occur. In order to achieve the second step, we assume that contact breaking events separated by large enough distances in space are uncorrelated with each other, and therefore we may treat each of them as independent.

\subsection{Local contact breaking distributions~\label{sec:local distributions}}
 
We begin by deriving the distribution of strains $\epsilon$ or polydispersity $\eta$ at which a particular contact would break for the first time irrespective of any other contact breaking or structural changes in the system. As we focus on the limit of low polydispersity and strain, it is reasonable to assume that the local structure around this contact has not deviated significantly from the crystalline structure. Let us focus on the $j^{th}$ contact of the particle situated at $\vec{r}$. Each contact breaking condition is described by a hypersurface in the $N_d$-dimensional space of disorder variables $\{\zeta_i\}$. This surface partitions the hypercube representing all possible quenched disorder configurations into two, a region where the contact is broken and the other where it remains unbroken. As the quenched disorder variables $\{\zeta_i\}$ are uniformly distributed in the $N_d$-dimensional hypercube, the volume of the hypercube that is partitioned by this single hypersurface representing the contact breaking condition, yields the cumulative probability $F_L(\epsilon)$ for the particular (or local) contact to remain unbroken at the strain $\epsilon$. The cumulative probability of the strain at which the $j^{th}$ contact of the particle situated at $\vec{r}$ remains unbroken is therefore given by
\begin{equation}
\begin{aligned}
    &F_L(\epsilon,\eta)=\\
    &\int_{-\frac{1}{2}}^{\frac{1}{2}}\prod_{\vec{r} ''} d \zeta(\vec{r}'') \Theta \left(z(\epsilon,\eta)-\sum_{\vec{r} '} C_j(\vec{r},\vec{r} ')\zeta(\vec{r}')\right).
    \end{aligned}
    \label{eq_exact_volume_local}
\end{equation}
We note that since the theta functions appear inside the integral in Eq.~(\ref{eq_exact_volume_local}), the cumulative first contact breaking distribution in Eq.~(\ref{eq_exact_volume}) is in general, not the product of all the individual local distributions. However, as the contact breaking events are local, we can approximate the total volume as originating from independent volumes, which do not intersect within the parameter regime we are interested in. This assumption will not be valid as more contacts break in the system and in particular, near the transition to the amorphous state, we expect this to not be true. Indeed, we show in the next Section that such an uncorrelated assumption reproduces the global first contact breaking strain distribution remarkably well. The distribution of the strains corresponding to the violation of the contact breaking condition for a given contact ($\vec{r},j$) can be obtained by taking a derivative of Eq.~\eqref{cdfth} with respect to $\epsilon$ as 
\begin{equation}
P_L(\epsilon)=-\frac{dF_L(\epsilon)}{d\epsilon}.
\label{eq_prob_vol_local}
\end{equation}

\begin{figure}[t!]
\hspace{-1cm}
\includegraphics[width=1.00\linewidth]{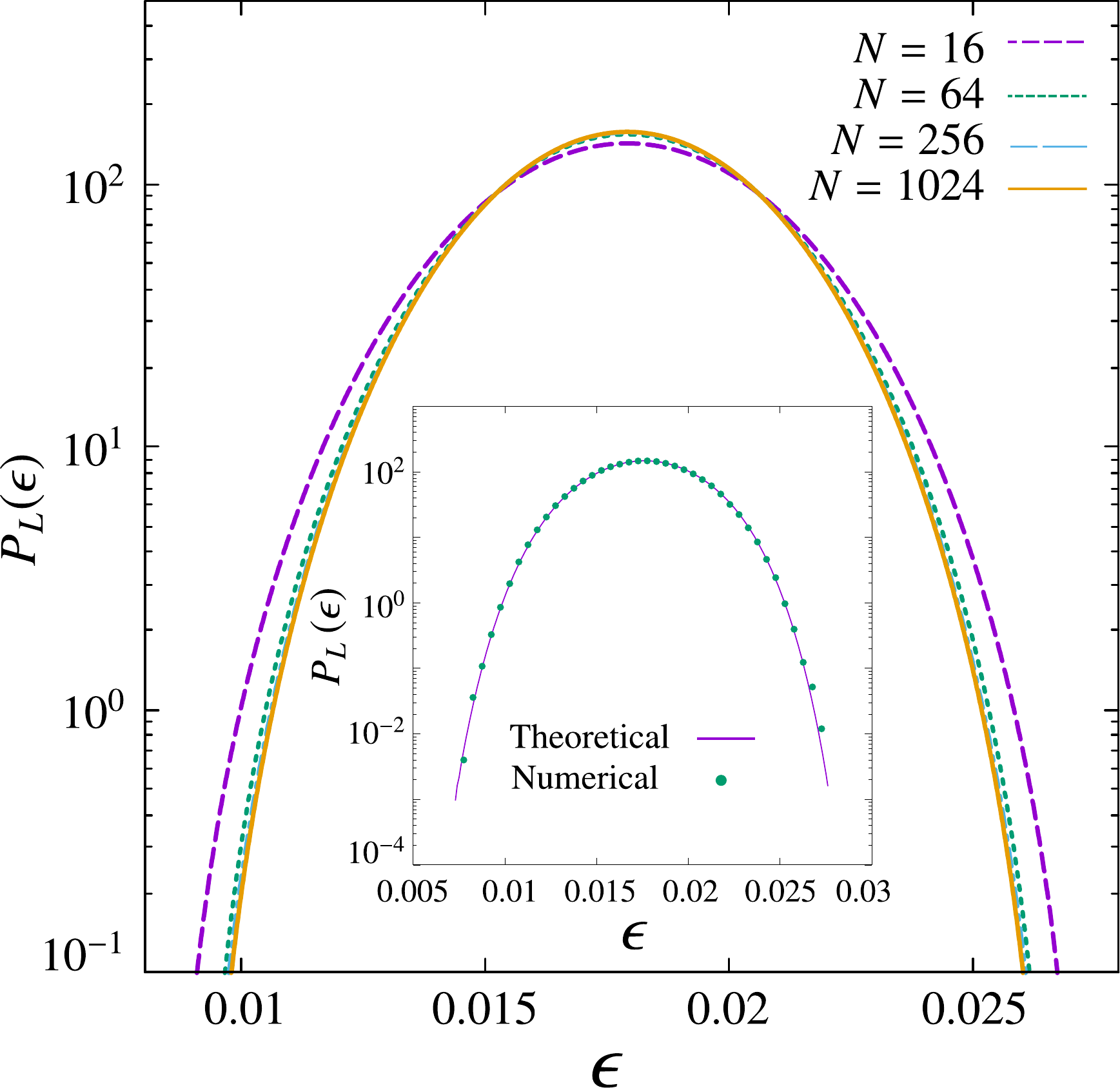}
\caption{Distribution of strains ($\epsilon$) at which the linearized contact breaking conditions given in Eq.~\eqref{eq_cb_condition} are violated for different system sizes ($N$). {\bf(Inset)} Comparison of the numerical results obtained from Monte-Carlo simulations of the individual contact breaking conditions and the theoretical predictions in Eqs.~\eqref{eq_prob_vol_local},~\eqref{cdfth} and~\eqref{cdfth2}. The results displayed are for a system size $N = 16$.}
\label{pthfig}
\end{figure}

We note that the distribution described in Eq.~(\ref{eq_exact_volume_local}) represents a generalized form of the Irwin-Hall distribution~\cite{marengo2017geometric,batsyn2013analytical}. Therefore, it is possible to compute the volume $(F_L)$ exactly (see Appendix~\ref{Irwin-Hall dist derivation} for details). This volume $(F_L)$ represents the probability that the particular contact mentioned in Eq.~\eqref{eq_cb_condition} is not broken at a particular value of $z$ (or corresponding strain  $\epsilon$) i.e. the Cumulative Distribution Function (CDF) of $z$. We have
\begin{equation}
F_L(z)=\frac{(z+2)^{N}-\sum_{k=1}^{N}(-1)^{k-1} g_k(z+2)}{N !\prod_{\vec{r}'}C_j(\vec{r},\vec{r}') },
\label{cdfth}
\end{equation}
where 
\begin{equation}
\begin{aligned}
g_k(z)=\sum_{1 \leq i_{1}<\cdots<i_{k} \leq N}  &\Theta\left(z-\sum_{n=1}^{k}C_j(\vec{r},\vec{r}_{i_n})\right) \\
&\times \left(z-\sum_{n=1}^{k}C_j(\vec{r},\vec{r}_{i_n})\right)^N.
\end{aligned}
\label{cdfth2}
\end{equation}
In the above equation, $i_n$ represents the particle index used in the partial sum involved in the function $g_k$. As the contact breaking condition of a single bond $(\vec{r},j)$ requires $N$ coefficients arising from all the particles, the function $g_k(z)$ conveniently groups all combinations involving $k$ coefficients. We note that the above expression is {\it exact}, and represents the cumulative distribution of the strain at which a particular contact breaking condition is violated. However computing the distributions using this form is computationally expensive for larger system sizes, and therefore it is more convenient to analyze this distribution in Fourier space as we detail in the next subsection. In the inset of Fig.~\ref{pthfig}, we display the match between the theoretical predictions and distributions obtained from a direct numerical simulation of the linearized contact breaking conditions for a small system size ($N=16$).

For large system sizes, utilizing the exact form involves the evaluation of large summations and it is therefore computationally expedient to obtain
$P_L(\epsilon)$ as an inverse Fourier transform of the generalized Irwin-Hall distribution in Fourier space $\tilde{P}_L(k)$. In this regard, the simpler expressions in Fourier space allow an easier determination of the local distributions. We have
\begin{equation}
\begin{aligned}
&\tilde{P}_L^{l}(k)=\exp\left[- \sum_{n=1}^{l}\alpha_n \left(\sum_{\vec{r}'}(C_j(\vec{r},\vec{r}',N))^{2n}\right) k^{2n} \right],
\end{aligned}
\label{eq_local_distribution}
\end{equation}
where the superscript $l$ denotes the number of terms summed in the series representation. We can then extract the local contact breaking distributions as
\begin{eqnarray}
\nonumber
P_{L}^{l}(\epsilon)&=&\frac{1}{2\pi} \int_{-\infty}^{\infty} dk e^{ik \epsilon} \tilde{P}^{l}(k),
\\
P_{L}(\epsilon)&=&\lim_{l\to \infty}P_{L}^{l}(\epsilon).
\label{irwinHall}
\end{eqnarray}
We note that increasing the number of terms in the sum given in the series representation of $\tilde{P}_L(k)$ in Eq.~(\ref{eq_local_distribution}) improves the accuracy of results for the first contact breaking strain distribution.

\subsection{Uncorrelated underlying distributions}

Given the underlying distributions of the individual contact breaking events, we next determine the distribution of the first contact breaking event, which is amenable to techniques of extreme value statistics of uncorrelated variables~\cite{gumbel1958statistics,david2004order}. 
Crucially, all the $3N$ contact breaking conditions arise from the $N$ independent variables $\{ \delta \sigma_i\}$. This suggests that only $N$ of the conditions are linearly independent, with the conditions for the other $2N$ contacts being dependent. This discrepancy can be resolved by noticing that the bonds of the system can be classified into three unique sets, along the $0$, $1$ and $2$ directions respectively (see Fig.~\ref{fig_schematic2}). For large system sizes, this classification splits the $N_d$-dimensional hypercube representing the quenched disorder at every site into three distinct classes. Each configuration of the disorder $\{ \delta \sigma_i \}$ can be uniquely identified with two rotated configurations $\{ \delta \sigma^{'}_i \} = \mathbb{R}_{\pi/3} \{ \delta \sigma_i \}$ and $\{ \delta \sigma^{''}_i \} = \mathbb{R}_{2\pi/3} \{ \delta \sigma_i \}$ that produce the same displacement fields, rotated by $\pi/3$ and $2\pi/3$ respectively. Therefore the statistics of the bond types $(0,1,2)$ are exactly identical. Moreover, for any distribution that requires an averaging over {\it all} configurations, only a single bond type contributes, with the other two being completely determined. This can be seen in the following example: if the probability that a first contact breaking event occurs at a given strain $\epsilon$ for the $0$ bonds $P(\epsilon|0)$ is determined, this leads to exactly the same probability for the $1$ and $2$ bonds as well i.e. for a bond breaking event at a given $\epsilon$, we have
\begin{eqnarray}
\nonumber
&&P(\epsilon|0) = P(\epsilon|1) = P(\epsilon|2),\\
&&P(\epsilon) = \frac{1}{3} \left( P(\epsilon|0) + P(\epsilon|1) + P(\epsilon|2) \right)= P(\epsilon|0).
\end{eqnarray}
Therefore, we may focus on a single bond type to determine the first contact breaking strain distribution, say along the $0$ direction. We next make an assumption of independence of the individual bonds in this class. This is reasonable as there exists a linear and invertible map between the bond lengths along the $0$ direction and the quenched disorder $\{ \delta \sigma \}$. We can therefore utilise the underlying marginal distributions for local contact breaking $P_{L}(\epsilon)$ derived in the previous Section.

\begin{figure}[t!]
\hspace*{-0.70 cm}
\includegraphics[width=1.0\linewidth]{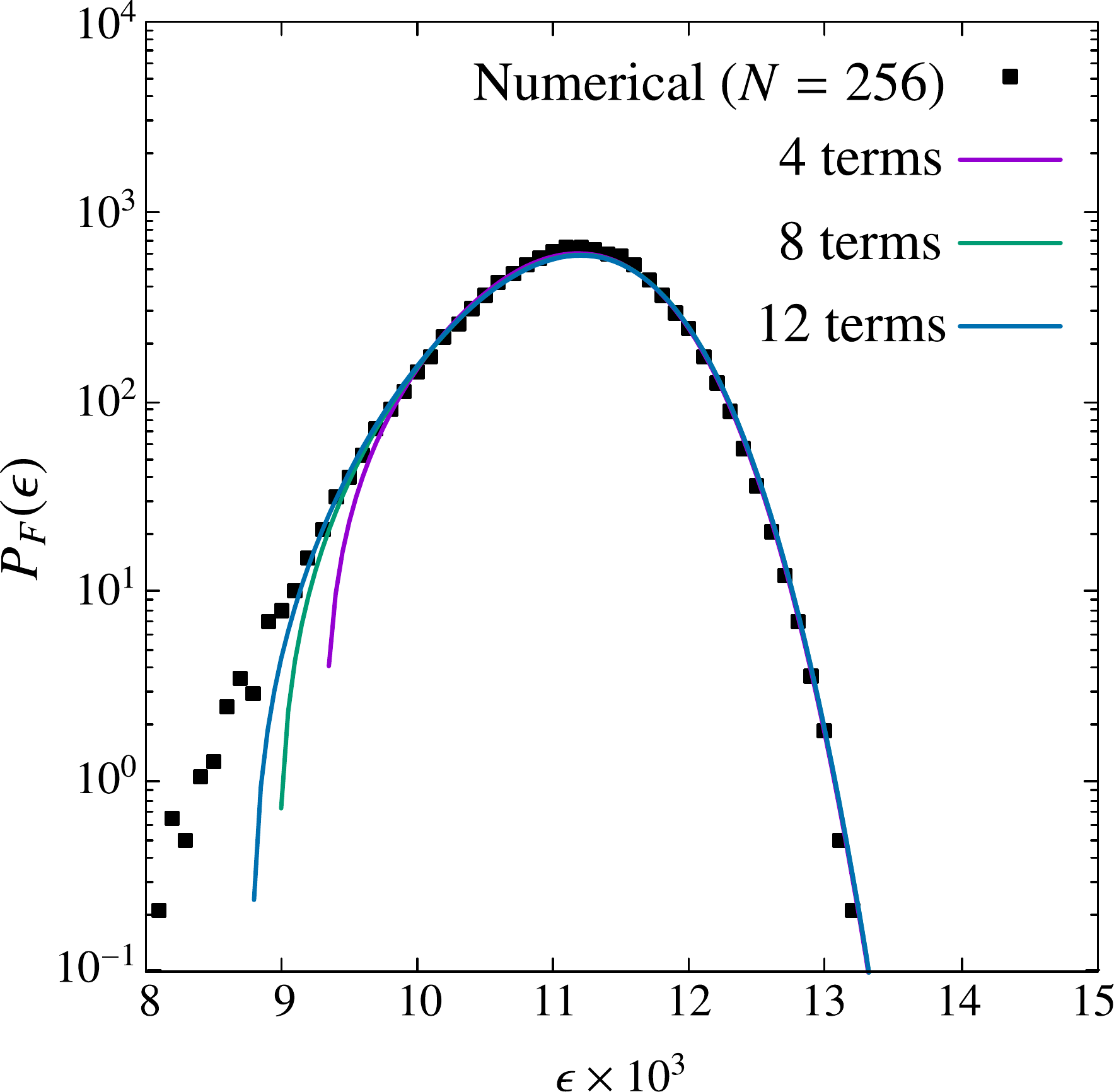}
\caption{First contact breaking strain distributions obtained from numerical simulations for a system size $N=256$ and polydispersity $\eta=0.015$. The solid lines represent the theoretical distribution obtained from Eq.~\eqref{eq_local_distribution} with an increasing number of terms in the Fourier transform, displaying the convergence of the theoretical predictions to the numerical results.
}
\label{NvsT}
\end{figure} 


Next, in order to derive the first contact breaking strain distribution, we analyze the strains at which each of the conditions for a given bond type (for example $j=0$) are violated. For a fixed quenched disorder $\{ \zeta_i \}$, these strains can be ordered as $\{\epsilon_1 < \epsilon_2 < \epsilon_3,... < \epsilon_N \}$. The minimum value in this set corresponds to the first contact breaking event in the system. Next, we make the assumption that these events are uncorrelated. This is reasonable as the contact breaking conditions for a given strain begin at different orthants of the hypercube representing the quenched disorder (see Fig.~\ref{fig_condition_surface}). 
Therefore, at smaller values of $\epsilon$, these surfaces do not intersect within the hypercube. As we show below, this uncorrelated assumption successfully predicts the first contact breaking strain distributions for a range of packing fractions and polydispersities. The joint probability distribution for the strains at which the local contact breaking events occur can then be expressed as
\begin{equation}
P(\{\epsilon_i\})=\prod_{i=1}^{N}P_{L}(\epsilon_i).
\end{equation}
Next, in order to theoretically derive the first contact breaking strain distribution, we make use of the analytic form for the local contact breaking distributions provided in Eq.~(\ref{cdfth}). 
The probability that the minimum value $ \text{Min} \{\epsilon_{i}\}$ is greater than $\epsilon$ can then be expressed as
\begin{equation}
\begin{aligned}
\mathrm  {Prob}(\mathrm {Min}\{\epsilon_{i}\}>\epsilon) = \prod_{i=1}^{N} \int_{\epsilon}^{\infty}d\epsilon_i P_{L}(\epsilon_i)\\
=\left(1-\int_{-\infty}^{\epsilon}dx P_{L}(x)\right)^N.
\end{aligned}
\label{cdf}
 \end{equation}

\begin{figure}[t!]
\hspace*{-0.70 cm}
\includegraphics[width=1.0\linewidth]{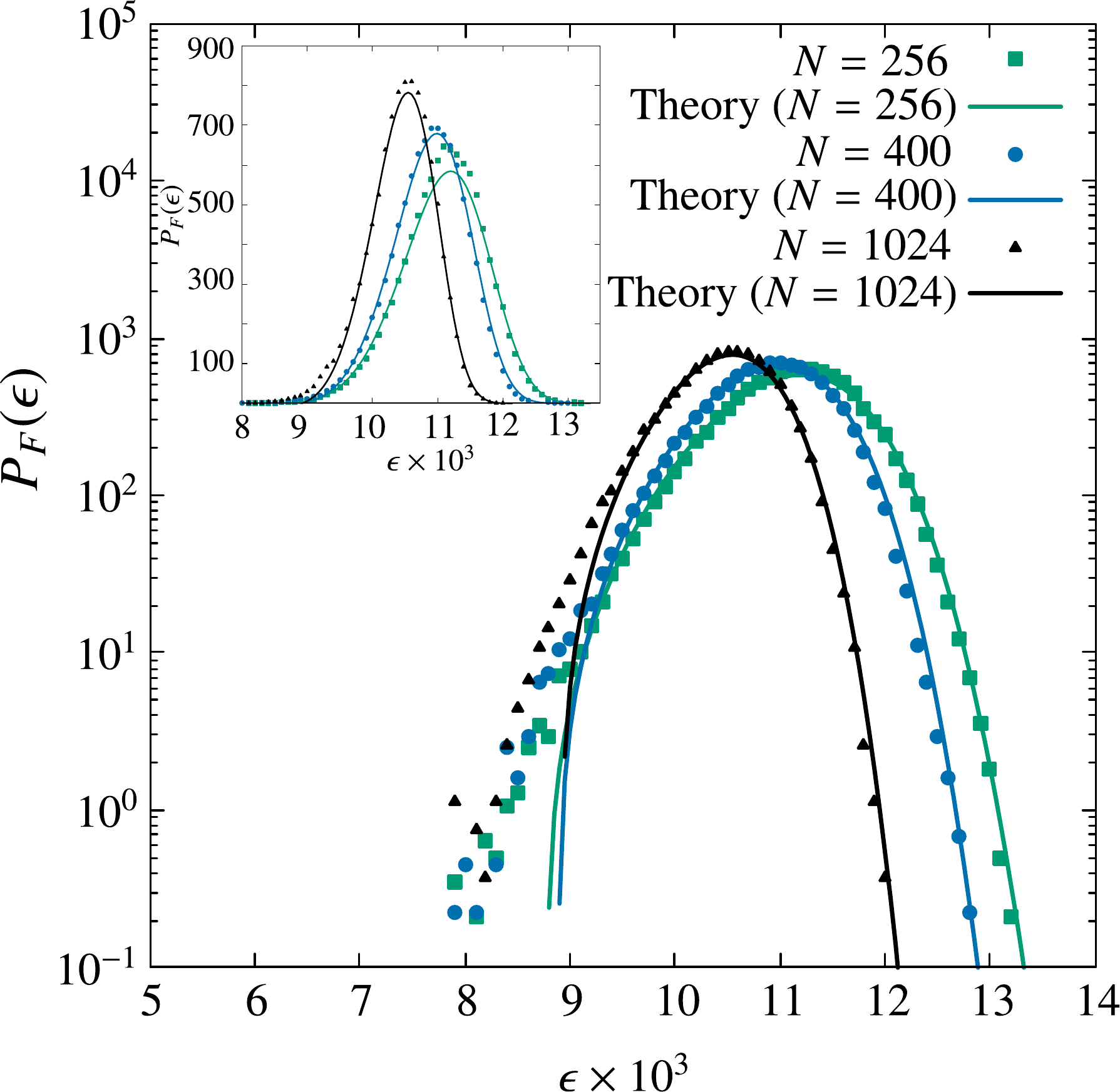}
\caption{Distribution of the first contact breaking strain $P_F(\epsilon)$ for different system sizes obtained from numerical simulations, along with the theoretical predictions from Eq.~(\ref{pdf}) using uncorrelated underlying distributions of local contact breaking strains $P_L(\epsilon)$.
{\bf(Inset)} The same distributions in linear scale.
}
\label{theory}
\end{figure} 

This represents the cumulative probability that no contact breaking occurs up to a volumetric strain of magnitude $\epsilon$. The first contact breaking strain distribution can therefore be obtained by taking a derivative of Eq.~\eqref{cdf} with respect to $\epsilon$ as,
\begin{equation}
\begin{aligned}
P_{F}(\epsilon)&=-\frac{d}{d\epsilon} \mathrm  {Prob}(\mathrm {Min}\{\epsilon_{i}\}>\epsilon)\\
&=N P_{L}(\epsilon)\left(1-\int_{-\infty}^{\epsilon}dx P_{L}(x)\right)^{N-1}.
\end{aligned}
\label{pdf}
\end{equation}




Therefore for small values of the strain, we obtain the scaling $P_F(\epsilon) \sim N P_{L}(\epsilon)$, which we also observe in our numerical simulations. In Fig.~\ref{NvsT}, we plot the first contact breaking distributions obtained from the underlying local distributions in Eq.~(\ref{eq_local_distribution}) with increasing number of terms ($l=4,8,12$). We find a remarkable match between the theoretical predictions and those obtained from direct numerical simulations, with no fitting parameters. Moreover, the results also depict the increasing agreement between theoretical and numerical results upon increasing the number of terms in the sum in Eq.~\eqref{eq_local_distribution}.

\section{System-Size Scaling~\label{sec:scaling}}

Finally, we analyze the behaviour of the first contact breaking strain distribution with increasing system size. We analyze the asymptotic behaviour of these distributions in the large system size limit. One of the important considerations in the stability of amorphous solids is the thermodynamic nature of the various phases associated with such materials. In this context it becomes important to study such systems in the limit with a large number of particles.
In order to extract the scaling behaviour of these distributions, we first analyze the scaling of the coefficients appearing in the Fourier space summation in Eq.~(\ref{eq_local_distribution}). We find
\begin{equation}
\sum_{i=1}^{N}(C_j(\vec{r},\vec{r}_i))^{n}=a_n +b_n/N.
\label{eq_summation}
\end{equation}
The scaling of these coefficients obtained by performing a numerical summation of the terms in Eq.~(\ref{eq_summation}) is displayed in Fig.~\ref{msum}. 
This scaling behavior indicates that for large system sizes, the coefficient $\sum_{i=1}^{N}(C_j(\vec{r},\vec{r}_i))^{n} \sim a_n$, implying that the local contact breaking distribution $P_L(\epsilon)$ becomes independent of system size. Using this limiting form of the underlying distributions, we compute the first contact breaking strain distribution using Eq.~(\ref{pdf}). 

\begin{figure}[t!]
\includegraphics[width=1.0\linewidth]{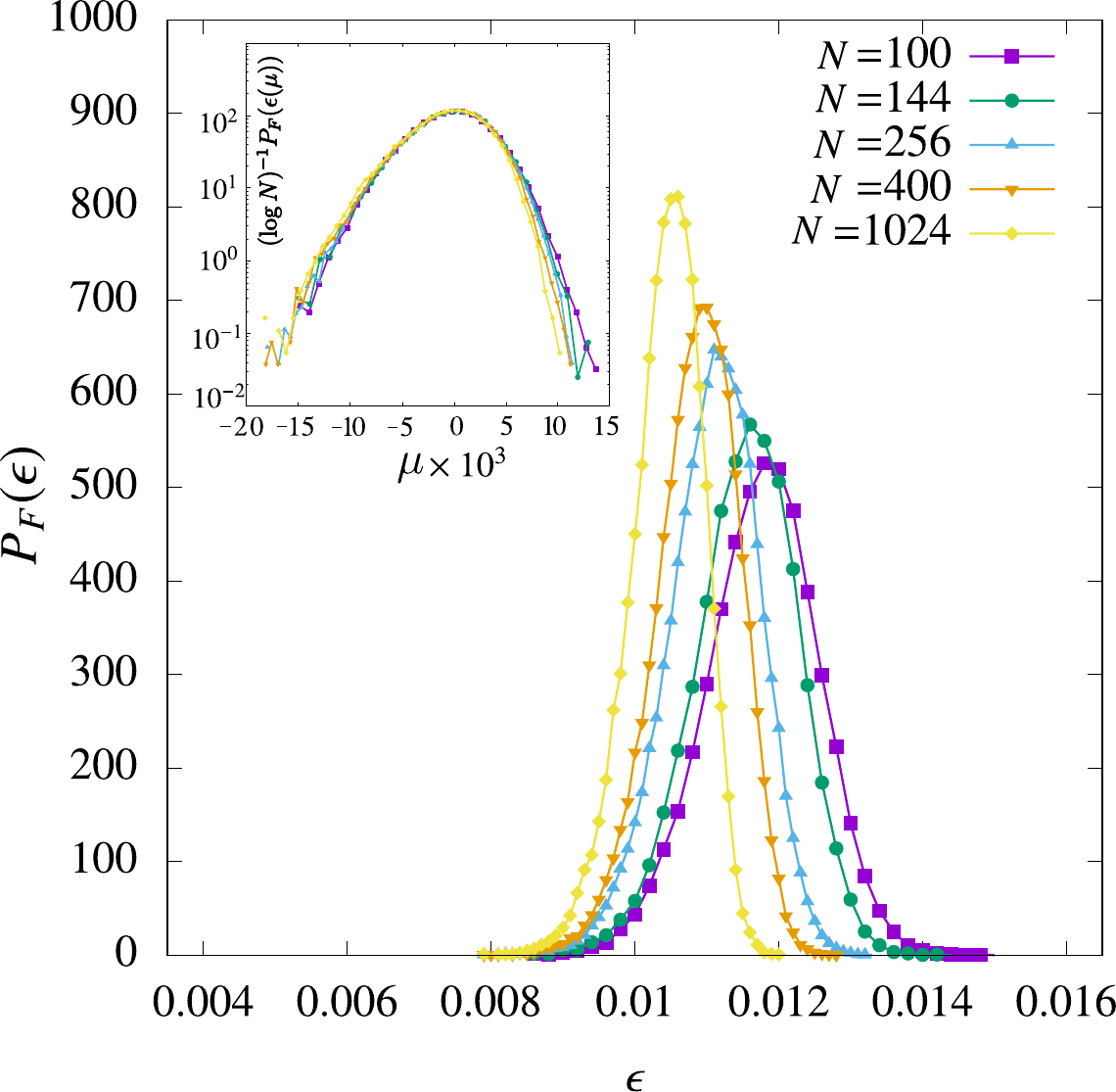}
\caption{
Distribution of first contact breaking strain at a polydispersity $\eta=0.015$ for different system sizes ($N$) obtained from numerical simulations. {\bf(Inset)} Scaling collapse of these distributions with the scaling variable $\mu=\alpha(N) \epsilon + \beta(N)$ with  $\alpha(N) = \log(N)$ and $\beta(N)=- c_1 + c_2 \log(N)$. Here $c_1 = 0.014$ and $c_2 = 0.0005$.
}
\label{fig_different_N}
\end{figure} 

In Fig.~\ref{fig_different_N}, we plot the first contact breaking distribution for different system sizes, ranging from $N=100$ to $N=1024$, for a fixed value of $\eta=0.015$. We find that these distributions display a good collapse with the scaling variable
\begin{equation}
    \mu =\alpha(N) \epsilon + \beta(N),
    \label{scaling_variable}
\end{equation}
where $\alpha(N) = \log(N)$ and $\beta(N)= -c_1 + c_2 \log(N)$, where $c_1$ and $c_2$ are constants that are independent of system size. Such a logarithmic scaling can be derived from the extreme value statistics of uncorrelated variables~\cite{gumbel1958statistics}. For an underlying distribution $p(x) \sim e^{-x^{\delta}}$, the distribution of the extreme value $x_{\text{min}} = \text{Min}\{ x_1, x_2, ..., x_N\}$ can be scaled with the variable $\mu = \alpha(N) x_{\text{min}} + \beta(N)$. The scaling of these quantities with the number of samples $N$, is given by $\alpha(N)\sim \left(\log(N)\right)^{1-\frac{1}{\delta}}$ and $\beta(N)\sim \log(N)$. 

For the case that we consider, the distribution $P_F(\epsilon)$ is obtained from the underlying distribution $P_L(\epsilon)$, which we have shown to be a generalized Irwin-Hall distribution. Since for this distribution, the tails decay much faster than an exponential, as is clear from the terms appearing in the Fourier transform in Eq.~(\ref{eq_local_distribution}) and the distribution plotted in Fig.~\ref{pthfig}. Therefore, in our case $\delta$ is a very large number, which leads to the scaling $\alpha(N) \sim \log(N)$. This scaling is displayed in the inset of Fig.~\ref{fig_different_N}, showing a very good scaling collapse. Additionally, as this represents an extreme value distribution arising from $N$ independent random variables drawn from an underlying distribution with a faster than power-law tail, it is well described by a Gumbel distribution. 

Previous studies have revealed that the average strain required to create a first plastic event in amorphous solids scales with the number of particles as $\langle \Delta \gamma \rangle \sim N^{\alpha}$ with $\alpha <0$~\cite{lerner2018protocol}.
Interestingly, this exponent decreases as the temperature is decreased. However, the value of the exponent in the zero temperature (athermal) limit is very small. Our study reveals that in the athermal limit, these distributions display a logarithmic scaling with system size. 




\section{Discussion}
\label{sec:Discussion}
In this paper, we have derived exact probability distributions for the strains at which the first stress drop events occur in disordered athermal crystals and verified our predictions with comparisons to numerical simulations. We demonstrated that the first stress drop event in this system coincides with the first contact breaking event and performed a detailed numerical as well as theoretical characterization of these events. This was made possible through an exact mapping of the computation of the cumulative distribution of strain to the computation of the volume of an $N_d$-dimensional convex polytope. For small numbers of defects $N_d$, we performed an exact numerical computation of these $N_d$-dimensional convex polytope volumes. We found a remarkable agreement between the distributions of strains where first stress drop events occur, generated using this exact volume computation and the strain distributions obtained from direct numerical simulations using athermal quasistatic strain. Finally, for large $N_d$, we derived the distribution of strains at which the first plastic failure occurs, assuming that individual contact breaking events are uncorrelated. We demonstrated that this accurately reproduces distributions obtained from direct numerical simulations.

Our study along with previous studies~\cite{acharya2020athermal,das2021long,acharya2021disorder} establishes the disordered crystal as a useful template system for understanding the mechanical properties of disordered athermal systems. This system enables us to compute exact theoretical results and understand mechanical properties of disordered athermal systems, which are usually modeled through coarse grained phenomenological descriptions~\cite{bi2015statistical}. The distribution of strains at which stress drop events occur has been of significant interest in several disordered systems, for example in the context of amorphous solids as it determines whether the material has a ``pseudogap'' with a non-zero  $\theta$ exponent~\cite{wyart2012marginal,lin2014density,muller2015marginal}. This distribution of stress drop events are known to be of crucial importance in determining the stability and yielding of amorphous solids~\cite{hentschel2015stochastic,karmakar2010statistical}. Our results at larger disorder, could provide a route towards understanding such properties in generic disordered amorphous materials. It is also straightforward to extend our techniques to understand the stability properties of such materials, subject to other mechanical perturbations such as shear. Our techniques can also be easily extended to near-crystalline structures in three dimensions. Finally, it would be interesting to extend the techniques developed in this paper, such as linearized trajectories in configuration space encountering hypersurfaces representing contact-breaking conditions \cite{sartor2021predicting}, to understand the stability properties of general amorphous packings.

\section*{Acknowledgments}
We thank Vishnu V.~Krishnan, Debankur Das, Pappu Acharya, Stephy Jose, Surajit Chakraborty, Soham Mukhopadhyay and Srikanth Sastry for useful discussions. JN was supported by NSF CBET award number 1916877. This project was funded by intramural funds at TIFR Hyderabad from the Department of Atomic Energy (DAE).

\appendix

\renewcommand\thefigure{\thesection\arabic{figure}}    
\section{Details of the Green's functions
\label{appendix:A}}
\setcounter{figure}{0}

\begin{figure*}[htp]
\centering
\includegraphics[width=0.900\linewidth]{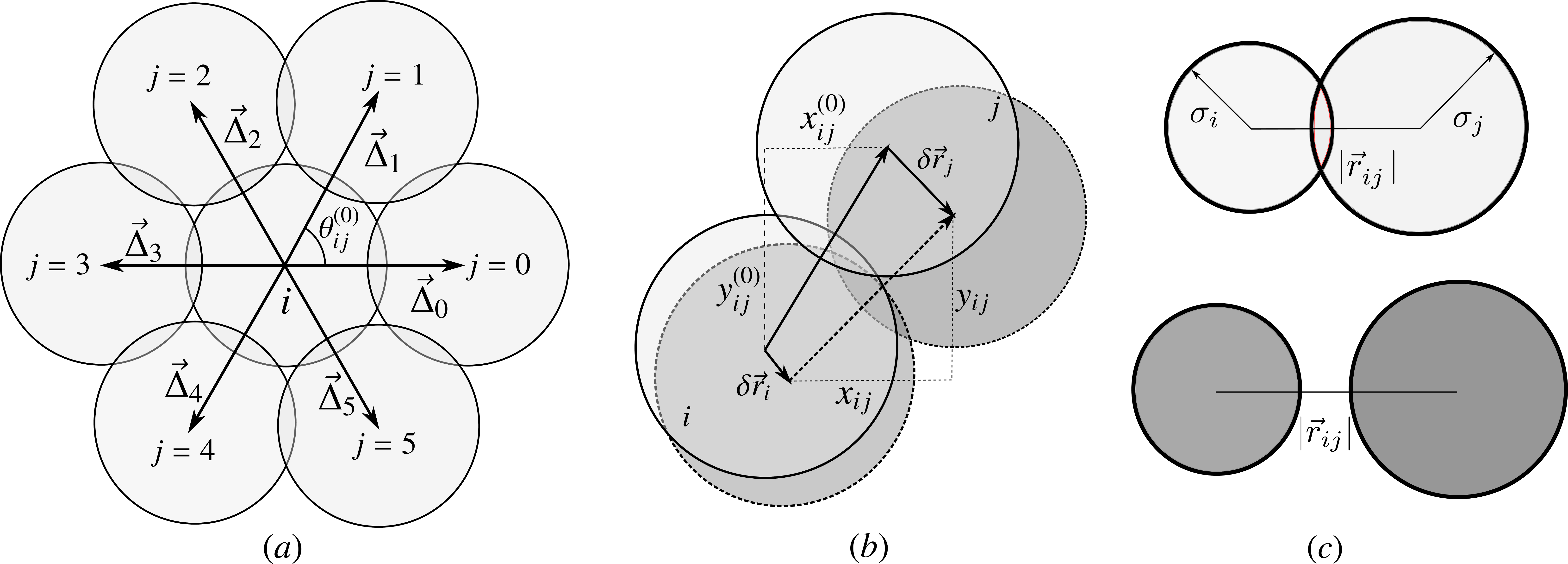}
\caption{
(a) A schematic representation of the neighbourhood of a single particle in the initial overcompressed triangular lattice configuration. (b) Relative displacements of two particles before and after energy minimization. (c) A schematic depiction of a contact breaking event. The particles move a finite distance away as the contact is broken and the system settles into a new energy minimum.}
\label{fig_schematic2}
\end{figure*}

Following Refs.~\cite{acharya2020athermal,acharya2021disorder,das2021displacement,das2021long}, we expand the force law in Eq.~(\ref{supp_force_law_equation}) to linear order and obtain the mechanical equilibrium condition on each grain as
\begin{equation}
\begin{aligned}
&\sum_{j=0}^{5} \left[\sum_{\nu}C_{i j}^{\mu \nu}\delta r^{\nu}_{ij} +C_{i j}^{\mu \sigma} \delta \sigma_{i j} \right]=0 ,
\end{aligned}   
\label{force balance linear}
\end{equation}
where $C_{i j}^{\mu \nu}$ and $C_{i j}^{\mu \sigma}$ are coefficients derived at linear order.
Due to the translation invariance of these coefficients, these equations can be solved exactly, yielding the displacement fields
\begin{equation}
\delta r^{\mu}(\vec{r}_{i})=\sum_{\vec{r}_n} G^{\mu}(\vec{r}_i-\vec{r}_n)\delta \sigma(\vec{r}_n).
\label{disp_field}
\end{equation}
The Green's functions in real space is given by $G^{\mu}(\vec{r}_i-\vec{r}_n)=\sum_{\vec{k}} e^ {-i \vec{k}. (\vec{r}_i-\vec{r}_n)}\tilde{G}^{\mu}(\vec{k})$, where $G^{\mu}(\vec{k})$ is the Green's function in Fourier space. These can be expressed as
\begin{equation}
    \tilde{G}^{\mu}(\vec{k})=\tilde{G}^{\mu x}(\vec{k})D^x(\vec{k})+\tilde{G}^{\mu y}(\vec{k})D^y(\vec{k}),
\end{equation}
with the individual components
\begin{eqnarray}
\begin{aligned}
\tilde{G}^{xx}(\vec{k}) = \Gamma_1 \frac{ \tilde{R}_0}{\Gamma_1\Gamma_2 -\Gamma^2_3} ,\\
\tilde{G}^{xy}(\vec{k}) = \Gamma_3 \frac{ \tilde{R}_0}{\Gamma_1\Gamma_2 -\Gamma^2_3}, \\
\tilde{G}^{yx}(\vec{k}) = \Gamma_3 \frac{ \tilde{R}_0}{\Gamma_1 \Gamma_2 -\Gamma^2_3}, \\
\tilde{G}^{yy}(\vec{k}) = \Gamma_2 \frac{ \tilde{R}_0}{\Gamma_1 \Gamma_2 -\Gamma^2_3}, \\
D^{\mu}(\vec{k})=\sum_{j=0}^{5}(1+e^{i \vec{k}.\vec{r}_{ij}^{(0)}})C^{\mu \sigma}_{ij}.
\label{eq_exact_green}
\end{aligned}
\end{eqnarray}
The functions $\Gamma_1$, $\Gamma_2$, $\Gamma_3$ have the following forms
\begin{eqnarray}
\begin{aligned}
\Gamma_1 =& (-1+4\tilde{R}_0 ) \cos (k_x) \cos (k_y)-2 (1-\tilde{R}_0)  \cos (2 k_x)\\
&-6 \tilde{R}_0 +3,\\
\Gamma_2 =& (-3+4 \tilde{R}_0) \cos (k_x) \cos (k_y)+2 \tilde{R}_0 \cos (2 k_x)\\
\nonumber
&-6 \tilde{R}_0 +3,\\
\Gamma_3 =&\sqrt{3}\sin(k_x)\sin(k_y).
\end{aligned}
\end{eqnarray}

\section{Derivation of contact breaking conditions~\label{appendixsec:contact breaking conditions}}

The condition for the contact between the $i^{th}$ and $j^{th}$ particle that are neighbours (see Fig.~\ref{fig_schematic2} (b)) to be not broken can be expressed as
\begin{equation}
|r_{i j}|^{2}<(\sigma_{i j})^{2},
\end{equation}
which can alternatively be expressed as
\begin{equation}
(x_{i j}^{(0)}+ \delta x_{i j})^2+( y_{i j}^{(0)} + \delta y_{i j})^2<(\sigma_{i j}^{(0)}+\delta\sigma_{i j})^2.
\end{equation}
Here $\{x_{ij}^{(0)},y_{ij}^{(0)}\}$ are the relative displacements between particles in the initial triangular lattice arrangement. Considering terms up to linear order in the above inequality, we arrive at
\begin{equation}
\begin{aligned}
R_0^2+\sum_{\mu=x,y}\left[ 2r_{ij}^{\mu(0)}\left(\delta r_{ij}^{\mu}\right)\right]<4\sigma_0^2+4\sigma_0\delta \sigma_{ij},
\end{aligned}
\label{condition_appendix}
\end{equation}
Next, we can express the relative distances between the particles in the initial triangular arrangement as
\begin{equation}
\begin{aligned}
&x_{ij}^{(0)}=R_0 \cos(\theta_{ij}^{(0)}),\\
&y_{ij}^{(0)}=R_0 \sin(\theta_{ij}^{(0)}),
\end{aligned}
\end{equation}
where $\theta_{ij}^{(0)}$ are the relative angles in the initial triangular arrangement (see Fig.~\ref{fig_schematic2} (a)). Next, the relative distance between nearest neighbours can be written using Eq.~\eqref{disp_field} as
\begin{equation}
\delta r^{\mu}_{ij}= \sum_{\vec{r}_n} \left[G^{x}(\vec{r}_j-\vec{r}_n)-G^{x}(\vec{r}_i-\vec{r}_n)\right]\delta \sigma(\vec{r}_n).
\end{equation}
Substituting these values of $\delta r^{\mu}_{ij}$ in Eq.~\eqref{condition_appendix} we arrive at
\begin{equation}
    \begin{aligned}
    \sum_{n=1}^{N}\left[\frac{R_0}{\sigma_0}\right. &\cos(\theta_{i j}^{0})(G^{x}_{jn}-G^{x}_{in})+\frac{R_0}{\sigma_0} \sin(\theta_{i j}^{0})(G^{y}_{jn}-G^{y}_{in})  \\
        &\left.-2( \delta_{in}+\delta_{jn} )\right]\delta \sigma(n) < \left(\frac{4\sigma_0^2-R_0^2}{2\sigma_0}\right).
    \end{aligned}
    \label{condition_appendix2}
\end{equation}
Here $G_{ij}^{\mu}=G^{\mu}(\vec{r}_i-\vec{r_j})$ and $\delta_{ij} \equiv \delta_{\vec{r}_i,\vec{r}_j}$. As the $i^{th}$ and $j^{th}$ particle are nearest neighbours, we can represent their positions more conveniently as
\begin{equation}
    \begin{aligned}
       &\vec{r}_n=\vec{r}',\\
        &\vec{r}_i=\vec{r},\\
        & \vec{r}_j=\vec{r}+\vec{\Delta}_j,
    \end{aligned}
\end{equation}
where $\vec{\Delta}_j$ represents the fundamental translation vectors of the triangular lattice (see Fig.~\ref{fig_schematic2}). We can then re-express Eq.~\eqref{condition_appendix2} using the above convention, which leads to Eq.~\eqref{coefficients} and Eq.~\eqref{coefficients2} in the main text.

\section{Threshold strain and polydispersity\label{threshold polydispersity}}
The $3N$ set of linear inequalities given in Eq.~\eqref{eq_cb_condition}, combined with the range of the disorder variables $\zeta(\vec{r})$ determines the boundary of the $N_d$-dimensional phase space of quenched disorder within which no contact is broken. A schematic diagram of this region for the case with disorder in two particles is represented in Fig.~\ref{fig_condition_surface}. The boundary of this region represents the locus of configurations where a contact breaking event occurs for the first time. Changing the value of $\eta$ or $\epsilon$ changes the boundary conditions provided in Eq.~\eqref{eq_cb_condition}, which in turn change the volume enclosed by these surfaces. All the surfaces defined in Eq.~\eqref{eq_cb_condition} first enter the hypercube representing the phase space of quenched disorder, when they intersect the corners of the hypercube. These corners are determined by the coordinates
\begin{equation}
\{\zeta^c\}= \frac{1}{2}\{\pm 1,\pm 1,....,\pm 1,\pm 1\}.
\end{equation}
Before any contact is broken, the surface given in Eq.~\eqref{eq_cb_condition} is closest to one of the corners of the hypercube. The coordinates of the closest corner corresponding to the contact breaking condition of the contact $(\vec{r},j)$ can be determined as
\begin{equation}
\{\zeta^c\}_{\vec{r},j}= \frac{1}{2}\{\sgn{C_j(\vec{r},\vec{r}_1)},\sgn{C_j(\vec{r},\vec{r}_2)}...,\sgn{C_j(\vec{r},\vec{r}_N)}\}.
\end{equation}
We next use this identification to determine the strain at which the first contact breaking event occurs, using Eq.~(\ref{eq_cb_condition}). We have
\begin{equation}
\frac{1}{2} \sum_{\vec{r}'} \sgn (C_{j}(\vec{r},\vec{r}'))C_{j}(\vec{r},\vec{r}')=  \frac{1}{2} \sum_{\vec{r}'}\left|C_{j}(\vec{r},\vec{r}')\right|=z(\eta,\epsilon^*).
\end{equation}
Therefore the solution to this equation determines the threshold value of strain $\epsilon^*$ before which no contact is broken in the system for a fixed value of $\eta$. Next, as the coefficients are exactly known, we can determine their scaling numerically. We find
\begin{equation}
\frac{1}{2} \sum_{\vec{r}'} \left|C_{j}(\vec{r},\vec{r}')\right|= z(\eta,\epsilon^*) = a+ b\log N,
\end{equation}
where $a$ and $b$ are constants that are independent of system size.
Since the underlying distributions of $\zeta$ have a finite support, the minimum value of the strain $\epsilon^*$ for any contact to break for the first time can be obtained from the above equation. We have
 \begin{equation}
     \epsilon^*  
     =\left(\frac{1-\tilde{R}_0^2}{1+\tilde{R}_0^2}\right) - \frac{ \eta(a+b \log N) }{2\left(1+\tilde{R}_0^2\right)}.
 \end{equation}
Alternatively, we can also determine the threshold value of polydispersity below which no contact is broken at zero strain (i.e. $\epsilon=0$). We have
\begin{equation}
\eta^* = \frac{4-(R_0/\sigma_0)^{2}}{a+b \log N}.
\end{equation}
For a system of $N=256$ particles, we find $\eta^* \sim 0.0127$. In our numerical simulations, the range of polydispersities considered are larger than this value. Therefore it is possible to encounter configurations at zero strain with contacts broken in the system. However these events are rare, and we do not observe them in the finite number of samples we consider.

\section{Derivation of generalized Irwin-Hall distribution\label{Irwin-Hall dist derivation}}

In this Appendix, we present a derivation of the generalized Irwin-Hall distribution, which we use to derive the distribution of strains at which the local contact breaking conditions in Eq.~\eqref{coefficients} are violated. Let $x_i$ for $i=1,2 ... N$ be independent random variables, each being uniformly distributed in the interval $[a_i,b_i]$. We are interested in the distribution of a linear combination of these variables $z=\sum_{i=1}^{N}\gamma_ix_i$. It is convenient to make a change of variables
\begin{equation}
X_i=\frac{x_i-a_i}{b_i-a_i},
\end{equation}
where the variables $X_i$ are distributed in the interval $[0,1]$. We next consider the rescaled variable
\begin{equation}
T = \sum_{i=1}^{N}\gamma_i X_i=z-\sum_{i=1}^{N}\frac{\gamma_i a_i}{b_i-a_i}.
\label{Tz_relation}
\end{equation}
This variable has the following exact distribution~\cite{marengo2017geometric,batsyn2013analytical}
\begin{equation}
    P(T)=\frac{1}{(N-1) !\prod_{i=1}^{N}\gamma_i }\left[T^{N-1}-\sum_{s=1}^{N}\frac{(-1)^{s-1}}{N}
 \frac{d g_s(T)}{dT}\right],
 \label{irwin}
\end{equation}
where 
\begin{equation}
g_s(T)=\sum_{1 \leq j_{1}<j_{2}<\cdots<j_{s} \leq N} \left(f\left(T-\sum_{n=1}^{s}\gamma_{j_n}\right)\right)^N,
\end{equation}
and
\begin{eqnarray}
f(x) = x \Theta(x).
\end{eqnarray}
Finally, the distribution of the variable $z$ can be obtained using the above expression and Eq.~\eqref{Tz_relation}. The cumulative distribution $F(z) = \int_{-\infty}^{z} P(z') dz'$ is given by
\begin{small}
\begin{equation}
\begin{aligned}
     &F(z)=\frac{1}{N!\prod_{i=1}^{N}\gamma_i } \times\\
     &\left[\left(z-\sum_{i=1}^{N}\frac{\gamma_i a_i}{b_i-a_i}\right)^{N-1}-\sum_{s=1}^{N}(-1)^{s-1}
g_s\left(z-\sum_{i=1}^{N}\frac{\gamma_i a_i}{b_i-a_i}\right)\right]
\end{aligned}
\end{equation}
\end{small}
\begin{figure}[t!]
\hspace*{-0.50 cm}
\includegraphics[width=0.95\linewidth]{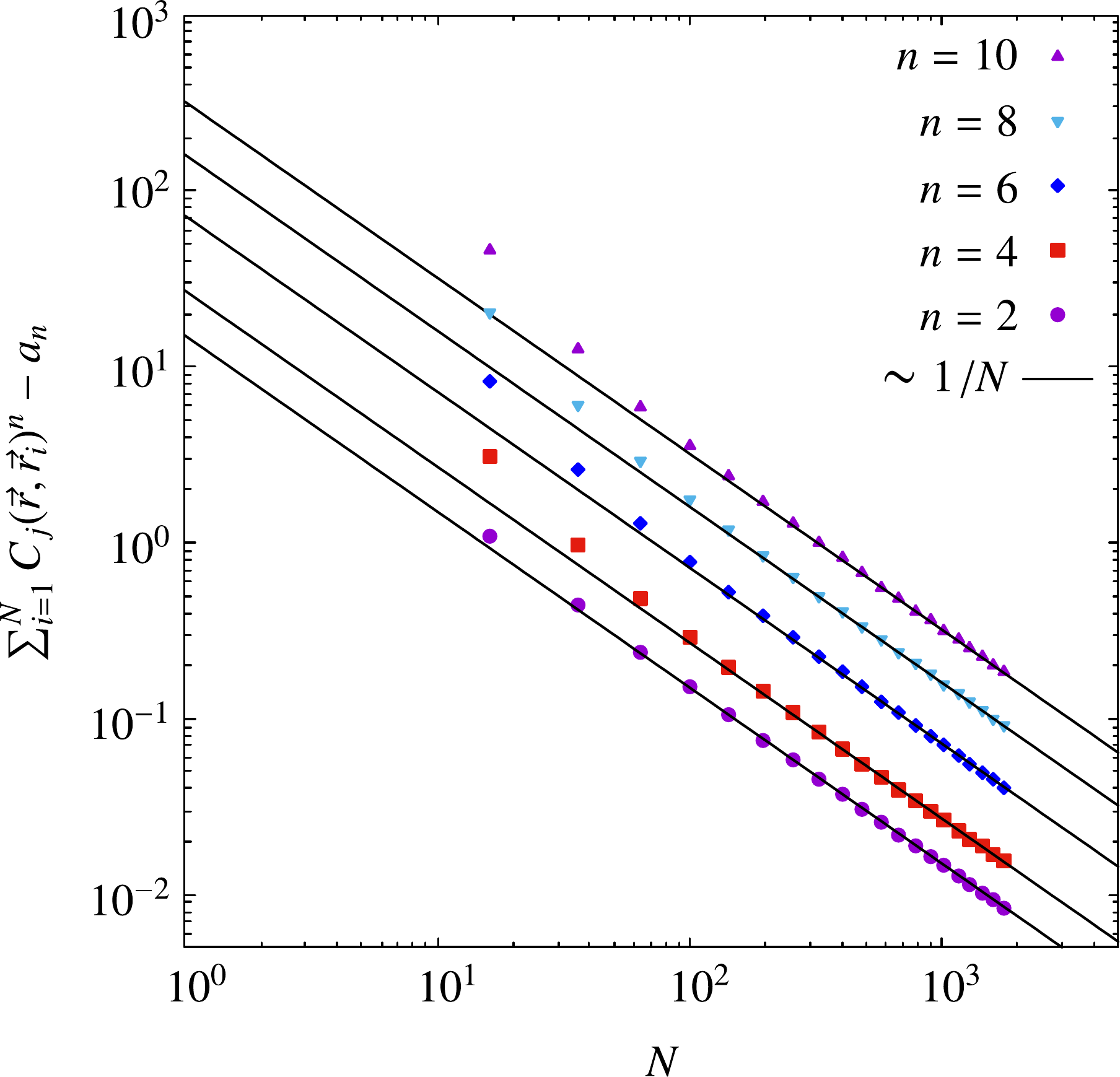}
\caption{
Variation of $\sum_{i=1}^{N}(C_j(\vec{r},\vec{r}_i))^n  $ with system size ($N$). Here $a_n=\lim_{N \to \infty}\sum_{i=1}^{N}(C_j(\vec{r},\vec{r}_i))^n$ only depend on the lattice structure and initial compression.
}
\label{msum}
\end{figure}
In the main text, we have used the above expression to compute the exact distributions of strain at which local contact breaking conditions are violated for small system sizes $(N =16)$, as shown in Fig.~\ref{pthfig}. 

We can alternatively derive the probability distribution of $z$ using the underlying distributions of $x_i$ as follows
\begin{equation}
\begin{aligned}
P(z)= &\int \prod_{i=1}^{N} dx_i  p(x_i)\delta(z-\sum_{i=1}^{N}\gamma_i x_i).
\end{aligned}
\end{equation}
We use the Fourier space representation of the delta function to simplify the above expression, we have
\begin{equation}
\begin{aligned}
P(z)=&\frac{1}{2\pi} \int_{-\infty}^{\infty} dk\int \prod_{i=1}^{N} d x_i  p(x_i) e^{ik\left(z-\sum_{i=1}^{N}\gamma_i x_i\right)}
\\
=&\frac{1}{2\pi} \int_{-\infty}^{\infty} dk e^{i k z}\prod_{i=1}^{N}\left[ \frac{\sin(k\gamma_i/2)}{k\gamma_i/2} \right]\\
&=\frac{1}{2\pi} \int_{-\infty}^{\infty} dk e^{i k z} \tilde{P}(k).
\end{aligned}
\end{equation}
We therefore obtain the Fourier transform of the required distribution
\begin{equation}
\begin{aligned}
    & \tilde{P}(k)=\prod_{i=1}^{N}\left[ \frac{\sin(k\gamma_i/2)}{k\gamma_i/2} \right].
\end{aligned}
\end{equation}

Next, taking a logarithm on both sides of above equation we arrive at
\begin{equation}
\begin{aligned}
&\log{\tilde{P}(k)}=\sum_{i=1}^{N}\log\left[ \frac{\sin(k\gamma_i/2)}{k\gamma_i/2} \right],\\
&=\sum_{i=1}^{N}\left[- \sum_{n=1}^{\infty}\alpha_n \gamma_i^{2n} k^{2n} \right]=- \sum_{n=1}^{\infty}\alpha_n \left[\sum_{i=1}^{N}\gamma_i^{2n}\right] k^{2n} ,
\end{aligned}
\end{equation}
where 
\begin{equation}
\begin{aligned}
    &\alpha_n = \frac{1}{2^{2n}}\left[\sum_{j,\text{mod}(n,j)=0}\frac{j}{n a_j^{n/j}} +\sum_{p,q,t,j}^{pt+qj=n} \frac{1}{a_t^p a_j^q}\right],
\end{aligned}
\end{equation}
and
\begin{equation}
a_j=(-1)^{j+1}(2j+1)!.
\end{equation}
The distribution in Fourier space is therefore given by
\begin{equation}
\tilde{P}(k)=\lim_{l\to\infty}\tilde{P}_l(k)=\lim_{l\to\infty}\exp\left[- \sum_{n=1}^{l}\alpha_n \left(\sum_{i=1}^{N}\gamma_i^{2n}\right) k^{2n} \right].
\end{equation}
Here $\tilde{P}_l(\vec{k})$ is the approximation of the Fourier transform considering $l$ terms in the exponential. 


\bibliography{First_Plastic_Drop_Bibliography}

\end{document}